\documentclass[fleqn,usenatbib]{mnras}

\usepackage{newtxtext,newtxmath}

\usepackage[T1]{fontenc}

\DeclareRobustCommand{\VAN}[3]{#2}
\let\VANthebibliography\thebibliography
\def\thebibliography{\DeclareRobustCommand{\VAN}[3]{##3}\VANthebibliography}

\usepackage[T1]{fontenc}
\usepackage{ae,aecompl}
\usepackage{amsmath}
\usepackage[dvipsnames]{xcolor}
\usepackage{graphicx}
\usepackage[colorinlistoftodos]{todonotes}
\usepackage{framed} 
\usepackage[framed]{ntheorem}
\usepackage{rotating}
\usepackage[titletoc]{appendix}
\usepackage{titlesec}
\usepackage{courier}       
\usepackage{titletoc}
\usepackage{rotating}
\usepackage{ntheorem}
\usepackage{esint}
\usepackage{empheq}
\usepackage{listings}
\usepackage{natbib}
\usepackage{pdflscape}
\usepackage{makecell}

\title[Stellar evolution impact on rotating star clusters]{The impact of stellar evolution on rotating star clusters: the gravothermal-gravogyro catastrophe and the formation of a bar of black holes}
\author[A. W. H. Kamlah et al. (2022)]{
	A. W. H. Kamlah$^{1,2}$\thanks{E-mail: kamlah@mpia-hd.mpg.de}, 
	R. Spurzem$^{2,3}$,
	P. Berczik$^{2,4,5}$,
	\newauthor
	M. Arca Sedda$^{2,6}$,
	F. Flammini Dotti$^{2,7,8}$,
	N. Neumayer$^{1}$,
	\newauthor
	X. Pang$^{7}$,
	Q. Shu$^{3,7}$,
	A. Tanikawa$^{9}$,
	and M. Giersz$^{10}$
	\\
	$^{1}$Max-Planck-Institut f\"ur Astronomie, K\"onigstuhl 17, 69117 Heidelberg, Germany \\
	$^{2}$Astronomisches Rechen-Institut, Zentrum f\"ur Astronomie, University of Heidelberg,
	M\"onchhofstrasse 12-14, 69120, Heidelberg, Germany \\
	$^{3}$Kavli Institute for Astronomy and Astrophysics, Peking University, Yiheyuan Lu 5, Haidian Qu, 100871, Beijing, China\\
	$^{4}$Konkoly Observatory, Research Centre for Astronomy and Earth Sciences, E\"otv\"os Lor\'and Research Network (ELKH), \\
	MTA Centre of Excellence, Konkoly Thege Mikl\'os \'ut 15-17, 1121 Budapest, Hungary \\
	$^{5}$Main Astronomical Observatory, National Academy of Sciences of Ukraine, 27 Akademika Zabolotnoho St., 03143 Kyiv, Ukraine \\
	$^{6}$Physics and Astronomy Department Galileo Galilei, University of Padova, Vicolo dell’Osservatorio 3, I–35122, Padova, Italy \\	
	$^{7}$Department of Physics, Xi’an Jiaotong-Liverpool University, 111 Ren’ai Rd., \\
	Suzhou Dushu Lake Science and Education Innovation District, Suzhou Industrial Park, Suzhou 215123, P.R. China \\
	$^{8}$Department of Mathematical Sciences, University of Liverpool, Liverpool L69 3BX, UK \\
	$^{9}$Department of Earth Science and Astronomy, College of Arts and Sciences, The University of Tokyo, 3-8-1 Komaba, Meguro-ku, Tokyo 153-8902, Japan \\
	$^{10}$Nicolaus Copernicus Astronomical Centre, Polish Academy of Sciences, ul. Bartycka 18, 00-716 Warsaw, Poland
}

\date{Accepted XXX. Received YYY; in original form ZZZ}

\pubyear{2022}

\begin{document}
	\label{firstpage}
	\pagerange{\pageref{firstpage}--\pageref{lastpage}}
	\maketitle
	
	\begin{abstract}
We present results from a suite of eight direct N-body simulations, performed with \textsc{Nbody6++GPU}, representing realistic models of rotating star clusters with up to $1.1\times 10^5$ stars. Our models feature primordial (hard) binaries, a continuous mass spectrum, differential rotation, and tidal mass loss induced by the overall gravitational field of the host galaxy. We explore the impact of rotation and stellar evolution on the star cluster dynamics. In all runs for rotating star clusters we detect a previously predicted mechanism: an initial phase of violent relaxation followed by the so-called gravogyro catastrophe. We find that the gravogyro catastrophe reaches a finite amplitude, which depends in strength on the level of the bulk rotation, and then levels off. After this phase the angular momentum is transferred from high-mass to low-mass particles in the cluster (both stars and compact objects). Simultaneously, the system becomes gravothermally unstable and collapses, thus undergoing the so-called gravothermal-gravogyro catastrophe. Comparing models with and without stellar evolution, we find an interesting difference. When stellar evolution is not considered, the whole process proceeds at a faster pace. The population of heavy objects tend to form a triaxial structure that rotates in the cluster centre. When stellar evolution is considered, we find that such a {\it rotating bar} is populated by stellar black holes and their progenitors. The triaxial structure becomes axisymmetric over time, but we also find that the models without stellar evolution suffer repeated gravogyro catastrophes as sufficient angular momentum and mass are removed by the tidal field.
	\end{abstract}
	
	\begin{keywords}
		methods: numerical – galaxies: star clusters: general – stars: general, black holes
	\end{keywords}
	
	
	
	\section{Introduction}
	\label{Section:Introduction}
	Present-day detectors and data processing methods have made it possible to resolve the photometry and kinematics of individual stars (even in components of binary and higher-order hierarchical stars) in star clusters \citep{Giesersetal2018,Giesersetal2019}. These observations reveal global bulk rotation of the star clusters and even resolve the rotational kinematics of the extremely dense star cluster cores. On top of this, the kinematic patterns of multiple populations in star clusters can and have been mapped out in numerous studies \citep{Bianchinietal2016,Bianchinietal2018,Bianchinietal2019,Ferraroetal2018,Lanzonietal2018a,Lanzonietal2018b,Kamannetal2016,Kamannetal2018a,Kamannetal2018b,Kamannetal2019,Sollimaetal2019,Tiongcoetal2019,Tiongcoetal2021}. Nowadays, we are also beginning to resolve the complex interaction between a star cluster and its tidal field and the imprint that the tidal field may leave on the internal cluster dynamics \citep{Tiongcoetal2016a,Tiongcoetal2016b,Tiongcoetal2017,Tiongcoetal2018}. \\
	With the use of these observations, we can refine existing theoretical models of star cluster dynamics. While supporting observational evidence of rotating and flattened star clusters accumulates, the majority of numerical and theoretical models of star clusters still rely on the simplistic assumption of spherical symmetry (e.g. \cite{Wangetal2016,Askaretal2017a,Rizzutoetal2021a,Rizzutoetal2021b,Kamlahetal2021}), which are supported by a wide range of models with fully self-consistent energy and angular momentum distribution functions (e.g. \cite{Plummer1911,King1962,Wilson1975}). Moreover, some methods simply require spherical symmetry. This is the case for Monte Carlo models and the mainstream Monte-Carlo codes are currently unable to evolve initially rotating star cluster models \citep{Henon1975,Cohn1979,Stodolkiewicz1982,Stodolkiewicz1986,Giersz1998,Gierszetal2015,Merritt2015,Askaretal2017a,Kremeretal2020a,Kremeretal2021}. Here, we briefly point out that \cite{Vasiliev2015} has developed a novel Monte Carlo method for simulating the dynamical evolution of stellar systems in arbitrary geometry. \\
	Recently, \cite{Lahenetal2020b} ran simulations of young massive star clusters forming in metal-poor starburst dwarf galaxies and found that the star clusters have significant angular momentum upon formation. In these simulations, the more massive star clusters tend to have larger angular momentum. But they also find that the angular momentum is not always aligned with flattening, thereby indicating a complex kinematic structure overall. Both observations and other simulations support these results and find that star clusters show significant fractality \citep{Balloneetal2020,Pangetal2021a}, and internal rotation at birth in general \citep{Balloneetal2021}. Velocity anisotropy has been observed in star clusters with detected elongated structures \citep{Pangetal2020,Pangetal2021a}, and these structures might be induced by rotation.\\
	\citet{AkiyamaSugimoto1989} already described the basic phenomena in a surprisingly small 1000 body direct $N$-body simulation; they found a four-phase star cluster evolution: ``(1) violent relaxation; (2) a gravogyro catastrophe of finite amplitude driven by the negative moment of inertia of a self-gravitating system through the transport of angular momentum; (3) a leveling off of the gravogyro instability where the transport of angular momentum is driven by coexisting, yet still slow, gravothermal instability; and (4) a relatively rapid gravothermal collapse'', directly cited from the abstract of \citet{AkiyamaSugimoto1989}. In the following years the focus shifted to the derivation of rotating equilibrium models, by \citet{Goodman1983a,LongarettiLagoute1996,VarriBertin2012}. These models are an extension of standard King models, adding a rotational parameter and a dependency of the distribution function on the angular momentum, and we denote them in the following as rotating King models. Such models were used as initial models for numerical solutions of the corresponding 2-D orbit-averaged Fokker-Planck (FP) equation. These models showed that not only the birth distribution, but also the long-term dynamical evolution of a star cluster is significantly affected by its initial bulk rotation, and follow-up work included binary heating and a stellar mass spectrum \citep{Kimetal2002,Kimetal2004,Kimetal2008,Fiestasetal2006}. Direct $N$-body models were resumed by \citet{Ernstetal2007,Hongetal2013}, in the first place to compare and check the numerical solutions of the FP equation. Rotation in nuclear star clusters was studied using the FP model \citep{FiestasSpurzem2010,Fiestasetal2012} and by $N$-body and semi-analytic models of  \citet{Szolgyenetal2018,Szolgyenetal2019,Szolgyenetal2021} - they were interested into the formation and evolution of rotating stellar or black hole disks in nuclear star clusters. Large and long term $N$-body simulations of star clusters, similar to globular clusters, were only recently published by \citet{Tiongcoetal2022,Livernoisetal2022}, though with some restrictions on the stellar mass function. \\
	In this paper we present and discuss the results of direct $N$-body simulations of rotating star clusters with and without stellar evolution. The models feature primordial (hard) binaries, a continuous mass spectrum, differential rotation, and tidal mass loss induced by the overall gravitational field of the host galaxy. \\
	The paper is structured as follows: in Sect.~\ref{Section:Gravothermal-gravogyro catastrophe}, we summarize the research status on the gravothermal-gravogyro catastrophe. In Sect.~\ref{Section:Methods} we discuss the methodology and in Sect.~\ref{Section:Initial conditions} we outline the initial conditions for the simulations. In Sect.~\ref{Section:Results} we present the simulation results and in Sect.~\ref{Section:Summary, conclusion and perspective} we summarize and conclude the work and we give a perspective on future work and open questions.
	
	\section{Gravothermal-gravogyro catastrophe}
	\label{Section:Gravothermal-gravogyro catastrophe}
	In the following we introduce the two main processes that mostly regulate the evolution of our rotating clusters, namely the gravothermal and gravogyro catastrophes. 
	
	\subsection{Gravothermal catastrophe}
	\label{Section:Gravothermal catastrophe}
	To understand the gravogyro catastrophe, it is didactically sensible to first illustrate the gravothermal catastrophe. It has been known that adding energy to a star cluster will make it cool down and expand \citep{LyndenBell1999}. This process was first proposed by \cite{Antonov1960,Antonov1961,Antonov1962}. He found that an isothermal gas sphere is the most probable state (maximum entropy $S$) of an initially spherical self-gravitating system of $N$ particles with energy $E$. However, he additionally found that this is not a global maximum. Below a certain density contrast between the central density $\rho_{c}$ (sphere of radius $r_{c}$) and the density at the edge of the sphere (sphere of radius $r_{e}$) $\rho_{e}$ ($\rho_{e}/\rho_{c}<1/709$), he showed that there exists no global maximum to the entropy $S$ at any fixed energy $E$. This effect is purely gravitational in nature and disappears in the absence of gravity \citep{LyndenBell1999}. \\
	\cite{LyndenBellWood1968} then developed the thermodynamic theory of self-gravitating gas spheres. Using linear response theories, they were able to demonstrate that for certain configurations of such systems, there exists no equilibrium state. Furthermore, they showed that the specific heat capacity $C_{\mathrm{V}}=\mathrm{d}E/\mathrm{d}T$ of the system becomes infinitely negative at around $3/100\lesssim\rho_{e}/\rho_{c}$ and approaches and ultimately reaches zero when the density contrast limit predicted by \cite{Antonov1962} is reached. Systems of self-gravitating gas spheres between the two limits are stable at fixed $E$ and $r_{\mathrm{E}}$ and they possess a negative heat capacity $C_{\mathrm{V}}$. For larger density contrasts than $1/709$, the system is unstable (no maximum entropy $S$). \\
	The following thought experiment is adapted from \cite{LyndenBell1999}. We can consider an isothermal gas sphere in a density contrast that eventually results in a negative specific heat capacity $C_{\mathrm{V}}$ as an analogy to a star cluster in order to understand the gravothermal catastrophe. We assume that the gas sphere expands adiabatically. We would observe a gas sphere with a much denser core than halo. As a result, mostly only the gas in the halo of the sphere will adiabatically expand. Consequently, the drop in temperature by the gas in the halo occurs much faster than the drop in temperature of the gas in the core. Keep in mind, that the specific heat capacity of the total system, $C_{\mathrm{V,total}}$ can be split up into the specific heat capacity of the gas in the core $C_{\mathrm{V,core}}$ core and the specific heat capacity of the gas in the halo $C_{\mathrm{V,halo}}$ Due to the resulting temperature gradient, heat will flow from the core to the halo of the gas sphere. As a result of the negative $C_{\mathrm{V,core}}$ the core will then contract and become hotter. The gas in the halo will also get hotter but in contrast to the gas in the core, it expands, because it has a positive $C_{\mathrm{V,halo}}$. If $C_{\mathrm{V,halo}}$ is very large, then this process will proceed indefinitely (in theory). The core will continuously lose more and more heat and this will cause it to contract further and further. This process is called the gravothermal catastrophe under the condition that $C_{\mathrm{V,total}}$ should continuously increase and reach zero once the boundary condition by \cite{Antonov1962} ($\rho_{e}/\rho_{c}=1/709$) is met. \\
	We now understand what happens in an adiabatically expanding, self-gravitating isothermal gas sphere. But in the context of stellar dynamics and realistic star clusters, the situation is much more complex. When replacing the gas molecules with actual stars in the thought experiment above, we now deal with a isothermal, self-gravitating sphere of stars. Heat is exchanged by repeated gravitational two-body encounters between the stars. The timescale for these encounters is much shorter at the centre of the cluster than at the outskirts of the star cluster. Therefore, when a star cluster adiabatically expands it is subject to the gravothermal catastrophe. The stellar density at the centre and the temperature (velocity dispersion) increases at ever smaller scales while the density in the halo decreases. This collapse would produce extremely large stellar densities at the core of the star cluster \citep{LyndenBellEggleton1980,InagakiLyndenBell1983,HachisuSugimoto1978,Hachisuetal1978,LyndenBell1999}. Therefore, we have to answer why we do not observe star clusters with such density profiles in the universe. Nowadays, we know that it stems from the fact that binary stars (and hierarchical systems) act as gravitational energy sources \citep{Aarseth1972,Aarseth1985,Heggie1975,Henon1975,Heggie1984} that can halt core-collapse. It has been shown that a collisional stellar system will evolve to a state of stars with predominantly radial orbits in the halo and a central core, which has an isotropic velocity distribution and possesses a central density that increases steadily \citep{Larson1970a,Larson1970b,Henon1972a,Henon1972b,Cohn1980,Bettwieser1983,BettwieserSpurzem1986}. The inclusion of binary stars on the other hand has a drastic effect, see \citet{BettwieserSugimoto1984}. They confirmed that binary formation happens near the centre of the star cluster and that they release energy. This effect causes the core to expand and to cool in temperature. The energy exchange between the core and the halo will result in an isothermal system. As a result, the gravothermal collapse occurs once more. This process may repeat many times in a simulation in the presence of binary stars \citep{LyndenBell1999}. \\
	So far, only closed-off systems were considered. If stars are allowed to escape the system by a series of weak gravitational encounters, a strong encounter or stellar evolution natal kicks, then this will accelerate the process of the gravothermal catastrophe and, ultimately, the whole system will disperse leaving behind only a single or a collection of extremely hard binary stars \citep{Padmanabhan1990}.
	
	\subsection{Gravogyro catastrophe and its coupling with the gravothermal catastrophe}
	\label{Gravogyro catastrophe and its coupling with the gravothermal catastrophe}
	The linear response theories developed by \cite{LyndenBellWood1968} were first applied to rigidly rotating and isothermal self-gravitating gas cylinders by \cite{InagakiHachisu1978}. They were able to define certain stability criteria for such systems, but were unable to define the coupling of the heat to angular momentum transport. To shed more light on this issue, \cite{Hachisu1979} used the theories by \cite{HachisuSugimoto1978} and he demonstrated that an unstable system as set up above has a negative specific moment of inertia even though its specific heat capacity $C_{\mathrm{V}}$ is positive (gravothermally stable). This can be explained by visualising a fluid element in a rigidly rotating and self-gravitating isothermal gas cylinder. When angular momentum is removed from the fluid element, then its angular speed also decreases. The region contracts towards the rotation axis of the gas cylinder. The moment of inertia of this fluid element decreases as a result. If the decrease of moment of inertia or the degree of contraction are large enough, then the angular speed actually becomes greater than its value before the removal of angular momentum. Ignoring gravity this may be coined as an effective negative specific moment of inertia in analogy to the negative specific heat capacity of gravothermal systems (see Sect.~\ref{Section:Gravothermal catastrophe}). Along these lines, \cite{Hachisu1979} called the underlying process the \textit{gravogyro catastrophe} in analogy to the gravothermal catastrophe discussed above (the angular velocity $\omega$ and the specific angular momentum $j$ correspond to the temperature $T$ and the specific entropy $s$ \citep{AkiyamaSugimoto1989}). \cite{Hachisu1979} then predicted two further important effects. Firstly, the gravogyro catastrophe cannot proceed indefinitely since the contraction of the star cluster is halted by binary stars. Secondly, the heat transport from the inner regions of the star cluster to outer regions assists the gravogyro catastrophe, because a loss of heat is also associated with a loss of pressure from the fluid element and thus its contraction is accelerated. \\	
	Later, the theories by \cite{LyndenBellWood1968,InagakiHachisu1978,Hachisu1979} were also applied to rotating and self-gravitating, isothermal gaseous disks and expanded to general three-dimensional bodies by \cite{Hachisu1982}, who confirmed that the instabilities are originating from a coupling of the gravothermal and the gravogyro catastrophes. They found that in general configurations of rotating bodies, both the gravothermal and the gravogyro catastrophes will prevail if either one of the following conditions hold: the central concentration of the gas needs to be large enough or if both the rotation is fast and the temperature of the gas is low enough. \cite{AkiyamaSugimoto1989} conducted first direct $N$-body simulations ($N=1000$, which is very small for statistical purposes \citep{EinselSpurzem1999}) using the direct $N$-body code \texttt{Nbody2} \citep{Aarseth1985}, which is a precursor to the direct $N$-body code \textsc{Nbody6++GPU} \citep{Wangetal2015,Wangetal2016} used in the work presented here. They observed a four-phase evolution in their simulations already outlined in the beginning of Sect.~\ref{Section:Introduction} and also concluded that such a series of evolutionary phases in combination with galactic tidal loss of stars would result in an overall loss of angular momentum from the cluster. 
	
	\subsection{2-D Fokker-Planck models vs. direct $N$-body simulations}
	Expanding on the solvers for the 2-D orbit-averaged Fokker-Planck (FP) equation in $(E,J_{\mathrm{z}})$ space developed by \cite{Goodman1983a}, \cite{EinselSpurzem1999} modelled the evolution of rotating stellar systems while assuming cylindrical coordinates and ignoring the existence of a third integral of motion. They propose a rotating King model in the form of 
	\begin{equation}
	\label{equation:rotatingKingmodel}
	f_{\mathrm{rk}} \propto \left(\mathrm{e}^{\beta E}-1\right)\times \mathrm{e}^{-\beta\Omega_0 J_{\mathrm{z}}}
	\end{equation}
	as a background distribution for the stars following \cite{LuptonGunn1987}, where $\beta = 1/(m\sigma_{\mathrm{c}}^2)$ and the dimensionless angular velocity is given by $\omega_{0}=\sqrt{9/4\times \pi Gn_{\mathrm{c}}}\times \Omega_{0}$. Potential-density pairs (see e.g. \cite{Binney2008}) for these models are created by relating $\beta$ to the King parameter $W_0$ via $W_0 = \beta m(\psi-\psi_{\mathrm{t}})$, where $\psi$ and $\psi_{\mathrm{t}}$ are the central King potential and the King potential at the truncation radius $r_{\mathrm{t}}$ as well as the number of stars and shells in the computation. \cite{EinselSpurzem1999} then established a family of rotating King models that are parameterised by pairs of $(W_{0},\omega_{0})$ using numerical and computational methods by \cite{Heyeyetal1959,Cohn1979,Spurzem1994,Spurzem1996}. \cite{EinselSpurzem1999} found that with increasing initial angular velocity parameter $\omega_{0}$, the system is driven into strong mass loss and it contracts moderately. Furthermore, the models exhibit the features for the gravogyro catastrophe found originally by \cite{Hachisu1979}: an increasingly faster rotating core, although angular momentum is transported outwards from the star cluster. \\
	The work by \cite{EinselSpurzem1999} was then improved through the inclusion of three-body binary heating \citep{Kimetal2002}. They performed simulations of equal-mass systems without stellar evolution or tides, but nevertheless they confirmed that the collapse time could be significantly reduced due to rotation. \cite{Kimetal2004} then improved the research further by including a two-component mass spectrum. Ultimately, they were able to show that generally the angular momentum is transported from the high mass to the low mass group as long as dynamical friction \citep{Chandrasekhar1943a,Chandrasekhar1943b,Chandrasekhar1943c,DosopoulouAntonini2017,Lingam2018} wins over the gravogyro catastrophe. In general, however, the underlying assumptions in the 2-D FP models by \cite{EinselSpurzem1999} (neglect of third integral of motion, axisymmetry, see also \cite{Spurzemetal2005} for a discussion of tidal fields) require comparisons with direct $N$-body simulations. For this purpose, \cite{Kimetal2008} then investigated single mass component models and showed that the FP results are generally consistent with the $N$-body calculations. Their results also confirmed earlier $N$-body simulations by \cite{Ernstetal2007}. The comparative studies between FP and direct $N$-body models were later expanded upon by \cite{Hongetal2013}, who showed that the cluster evolution is accelerated by not only the initial rotation but also the mass spectrum of the cluster. They also demonstrated that the total angular momentum and the total mass of the cluster both decrease rapidly, while a bar-like structure forms and persists in the cluster centre. The formation of a bar and its subsequent fairly rapid dissolution was already found earlier in the pioneer simulations by \citet{AkiyamaSugimoto1989}. Furthermore, it was confirmed that there is no conflict with observed limits of Galactic globular cluster rotation by expanding upon earlier comparisons between the FP models and observations from \cite{Fiestasetal2006,FiestasSpurzem2010}. \cite{Szolgyenetal2019}, who initialised their $N$-model simulations with rotating King models from \cite{LongarettiLagoute1996}, found a process of anisotropic segregation of heavy masses towards the central region, forming a disk-like structure. This has been proposed earlier for galactic nuclei \citep{Szolgyenetal2018} and studied in more detail in \cite{Szolgyenetal2021}. The formation of such a disk is very likely linked to the gravothermal-gravogyro catastrophe and similar to the formation of the bar-like structure found by \cite{AkiyamaSugimoto1989,Hongetal2013}. \\
	The work presented in this paper adds to the large body of theoretical work listed above. For the first time, we study the impact of initial bulk rotation, realistic stellar evolution mass loss models in combination with primordial binaries and stars drawn from a continuous IMF \citep{Kroupa2001} and the impact of the tidal field on the global dynamics of the star clusters. With these settings, we study the development, evolution and coupling of the gravothermal and gravogyro catastrophes using direct $N$-body methods during the pre- and post-core collapse phases of star cluster evolution over 1~Gyr.  
	
	\section{Methods}
	\label{Section:Methods}
	\subsection{\textsc{Nbody6++GPU}}
	The rotating star cluster models are evolved using the state-of-the-art direct force integration code \textsc{Nbody6++GPU}, which is optimised for high performance GPU-accelerated supercomputing \citep{Spurzem1999,NitadoriAarseth2012,Wangetal2015}. It is a successor to the many direct force integration $N$-body codes of gravitational $N$-body problems, which were originally written by Sverre Aarseth (\citet{Aarseth1985,Spurzem1999,Aarseth1999a,Aarseth1999b,Aarseth2003b,Aarseth2008} and sources therein). 
	\\
	The code is optimised for large-scale computing clusters by utilising MPI \citep{Spurzem1999}, SIMD, OpenMP and GPU \citep{NitadoriAarseth2012,Wangetal2015} parallelisation techniques. In combination with the Kustaanheimo-Stiefel (KS) regularisation \citep{Stiefel1965}, the Hermite scheme with hierarchical block time-steps \citep{McMillan1986,Hutetal1995,Makino1991,Makino1999} and the Ahmad-Cohen (AC) neighbour scheme \citep{AhmadCohen1973}, the code allows for star cluster simulations of realistic size without sacrificing astrophysical accuracy by not properly resolving close binary and/or higher-order subsystems of (degenerate) stars. With \textsc{Nbody6++GPU} we can include hard binaries and close encounters (binding energy comparable or larger than the thermal energy of surrounding stars) using two-body and chain regularization \citep{MikkolaTanikawa1999a,MikkolaTanikawa1999b,MikkolaAarseth1998}, which permits the treatment of binaries with periods of days in conjunction and multi-scale coupling with the cluster environment. The AC scheme permits for every star to divide the gravitational forces acting on it into the regular component, originating from distant stars, and an irregular part, originating from nearby stars (``neighbours''). Regular forces, efficiently accelerated on the GPU, are updated in larger regular time steps, while neighbour forces are much more fluctuating and need update in much shorter time intervals. Since neighbour numbers are usually small compared to the total particle number, their implementation on the CPU using OpenMP \citep{Wangetal2015} provides the best overall performance. Post-Newtonian dynamics of relativistic binaries is currently still using the orbit-averaged Peters \& Matthews formalism \citep{PetersMathews1963,Peters1964}, as described e.g. in  \citet{DiCarloetal2019,DiCarloetal2020a,DiCarloetal2020b,DiCarloetal2021,Rizzutoetal2021a,Rizzutoetal2021b,ArcaSeddaetal2021}.
	
	\subsection{\textsc{McLuster} \& \textsc{fopax}}
	\label{McLuster,fopax,rotinit}
	Our initial N-body particle distribution and velocities are obtained in three steps. \\
	Firstly, the star clusters are initialised with \textsc{McLuster} \citep{Kuepperetal2011a,Kamlahetal2021,Levequeetal2022}. This code is used to either set up initial conditions for $N$-body computations or to generate artificial star clusters for direct investigation \citep{Kuepperetal2011a}.  The \textsc{McLuster} output models can be read directly into the \textsc{Nbody6++GPU} as initial models (also other codes, e.g. \textsc{MOCCA} \citep{Kamlahetal2021}). This makes \textsc{McLuster} the perfect tool to initialise realistic star cluster simulations. The input parameters are given in the Section \label{Section:Initial conditions} and they can be found in Tab. \ref{Global_properties_newtidal}. \\
	Secondly, we generate 2-D Fokker-Planck initial models as used in \citet{EinselSpurzem1999,Kimetal2002,Kimetal2004,Kimetal2008} with the Fokker-Planck code named \textsc{fopax}. 
	The code produces a 2-D mesh based output of density $\rho$ and velocity dispersions $\sigma$ as a function of $r$ and $z$ based on the rotating King model $f(E, J_{\mathrm{z}})$ that are characterised by a pair of parameters $(W_{0},\omega_{0})$ (see Eq.~\ref{equation:rotatingKingmodel}). \\
	Thirdly, a Monte Carlo rejection technique is then used to generate a discrete system of $N$ particles following the known distributions of $\rho$ and $\sigma$. The output is in $N$-body format (one line per particle, mass, and 3-D position, velocity data). This $N$-body distribution is combined with the \textsc{McLuster} $N$-body distribution and all data is scaled to standard H\'enon units. As a result, we have an initial star cluster model that is a rotating King model $N$-body distribution with the chosen IMF and all relevant binary orbital parameter distributions conserved from \textsc{McLuster}.\\
	It is important here that the dimensionless King model parameter $W_0$ is identical in both \textsc{McLuster} and \textsc{fopax} (In our set-up $W_{0}=6.0$). In this way, we create models $(W_{0}=6.0,\omega_{0}\in[0.0, 0.6, 1.2, 1.8]$; see Tab.~1 in \cite{EinselSpurzem1999} for up to $\omega_{0}=1.0$) in the construction of the initially rotating $N$-body distributions of star cluster models presented in this paper. Models with $(W_{0}, \omega_{0})=(6.0, 0.0)$ are identical to traditional King models with $W_0 = 6.0$. \\
	Furthermore, the rotating King model initial distributions are initially more compact with increasing $\omega_{0}$ (see Fig. 1 in \cite{EinselSpurzem1999}). Therefore, the structural input parameters from \textsc{McLuster}, such as the half-mass radius $r_{\mathrm{h}}$, are (slightly) changed in this step. Since the traditional calculation of the half-mass radii $r_{\mathrm{h}}$ and by extension also the Lagrangian radii $r_{\mathrm{Lagr}}$ rely on the assumption of spherical symmetry, which breaks down for the rotating models (and in general, also for initially spherical star clusters in tidal fields), they can only be used as an approximate or indicative measure for the global, structural evolution of the star clusters. All of this also implies that the initial half-mass relaxation times are smaller for increasing $\omega_{0}$ (see Tab. 1 in \cite{EinselSpurzem1999}). 
	
	\section{Initial conditions}
	\label{Section:Initial conditions}
	\subsection{Star cluster parameters}
	\begin{table}
		\begin{tabular}{|l|l|}
			\hline \textbf{Quantity} & \textbf{Value} \\
			\hline \hline \textbf{Particle number} & $1.1\times 10^5$ \\
			\hline \textbf{Binary fraction} $f_{\text{b}}$ & $10.0\%$  \\
			\hline \textbf{Half mass radius} $r_{\text h}$ & $1.85$~pc  \\
			\hline \textbf{Tidal radius} $r_{\text{tid}}$ & $65.59$~pc   \\
			\hline \textbf{IMF} & \makecell{Kroupa IMF \citep{Kroupa2001} \\ ($0.08-150$)~$\text M_{\odot}$} \\
			\hline \textbf{Density model} & \makecell{King model \citep{King1962}\\ $W_0=6.0$} \\
			\hline \makecell{\textbf{Eccentricity} \\ \textbf{distribution} $f(e)$} & \makecell{Thermal ($f(e)\propto e^2$)} \\
			\hline \makecell{\textbf{Semi-major axis} \\ \textbf{distribution} $f(a)$} & \makecell{uniform in $\mathrm{log}(a)$ \\ between the sum of the radii \\ of the two binary stars and 100~AU } \\
			\hline \makecell{\textbf{mass ratio} \\ \textbf{distribution} $f(q)$} &  \makecell{uniform distribution of mass ratio \\ (0.1$<q<$1.0) for $m>5$~$\text{M}_{\odot}$ and \\random pairing for \\ the remaining binaries \\ \citep{Kiminkietal2012,SanaEvans2011} \\ \citep{Sanaetal2013a,Kobulnickyetal2014}.}\\
			\hline \hline
		\end{tabular}
		\caption{Initial parameters that are identical across all eight initial models for the \textsc{Nbody6++GPU} simulations.}
		\label{Initial_conditions}
	\end{table}	
	
	\begin{table}
		\begin{tabular}{|l|l|l|}
			\hline \textbf{Model ID} & \textbf{Stellar evolution?} & $\omega_0$ \\
			\hline \hline \texttt{SEV}$\omega_00.0$ & yes & $0.0$ \\
			\texttt{SEV}$\omega_00.6$ & yes & $0.6$ \\
			\texttt{SEV}$\omega_01.2$ & yes & $1.2$ \\
			\texttt{SEV}$\omega_01.8$ & yes & $1.8$ \\
			\hline \hline \texttt{noSEV}$\omega_00.0$ & no & $0.0$ \\
			\texttt{noSEV}$\omega_00.6$ & no & $0.6$ \\
			\texttt{noSEV}$\omega_01.2$ & no & $1.2$ \\
			\texttt{noSEV}$\omega_01.8$ & no & $1.8$ \\
			\hline \hline
		\end{tabular}
		\caption{Model identifiers (Model ID) for the eight \textsc{Nbody6++GPU} simulations.}
		\label{Model_IDs}
	\end{table}	
	
	The initial models from \textsc{McLuster} \citep{Kuepperetal2011a,Kamlahetal2021,Levequeetal2022} are constructed as smaller mock models of the Milky Way GC NGC3201 and are shown in Tab.~\ref{Initial_conditions}. The initial number of objects is set to $10^5$ with a binary fraction of $0.1$. This yields a total number of stars of $1.1\times 10^5$. Our clusters have an initial cluster mass of $6.41\times 10^4$~$\mathrm{M}_{\odot}$. As sketched out above, we use a King density model with a King model parameter of $W_0 = 6.0$ \citep{King1962}. The model shows no initial mass segregation and is unfractal \citep{GoodwinWhitworth2004}. The model is initially in virial equilibrium. The half-mass radius is set to $r_{\rm h}$=1.85~pc. As outlined in Sect.~\ref{McLuster,fopax,rotinit}, the initial model from \textsc{McLuster} is then redistributed with a rotating King model, which are more compact than their non-rotating counterparts \citep{EinselSpurzem1999}. Therefore, the internal structural parameters such as the $r_{\rm h}$ and $r_{\rm c}$ change in this initialisation step from their original \textsc{McLuster} $N$-body distribution (see already Fig.~\ref{Global_properties_newtidal}).\\
	We use a Kroupa IMF \citep{Kroupaetal2001} between 0.08~$\mathrm{M}_{\odot}$ and 150.0~$\mathrm{M}_{\odot}$. The binaries are paired in their mass ratios $q$ following \citep{Kiminkietal2012,SanaEvans2011,Sanaetal2013a,Kobulnickyetal2014}, meaning that we have a uniform distribution of mass ratios (0.1$<q<$1.0) for $m>5$~$\text{M}_{\odot}$ and random pairing for the remaining binaries. Their semi-major axes are distributed uniformly in log-scale between the sum of the radii of the two binary stars and 100~AU. The eccentricity distribution is thermal. \\
	The cluster's absolute metallicity is set to $Z=0.00051$. We put our cluster initial models on a circular orbit around the Galaxy of radius 13.3~kpc (according to \citep{Caietal2016} a circular orbit can be chosen such that the mass loss evolution of the cluster is similar compared to the eccentric orbit of NGC3201 (between 8.60 and 29.25~kpc, with eccentricity $e=$0.55 according to Gaia DR2 data \citep{Helmietal2020a})) around a point-mass MW of mass $1.78\times 10^{11}$~$\mathrm{M}_{\odot}$ (assuming a circular velocity $v_c = 240.0$~$\mathrm{kms}^{-1}$ at the Solar distance) \citep{Helmietal2020a,BobylevBajkova2020}. For our cluster models this yields an initial tidal radius of 65.59~pc. Therefore, the models are very tidally underfilling. \\
	In the interest of aiding the discussion, we introduce model IDs for our eight individual runs, see Tab.~\ref{Model_IDs}. For example, the non-rotating model without stellar evolution is named \texttt{noSEV}$\omega_00.0$, while the rotating model with $\omega_0=1.2$ and stellar evolution switched on is named \texttt{SEV}$\omega_01.2$. The details of the stellar evolution parameters are discussed below. Furthermore, we will refer to the group of models without stellar evolution as \texttt{noSEV} models and to the group of models with stellar evolution as \texttt{SEV} models from here on after. 
	
	\subsection{Stellar evolution parameters}
	We follow the \texttt{level C} stellar evolution as presented in \citet{Kamlahetal2021}, which also describes the stellar evolution routines and parameters in detail. We use the metallicity-dependent winds following \citet{Vinketal2001,VinkdeKoter2002,VinkdeKoter2005,Belczynskietal2010} across the full mass range. For the compact object evolution, we use remnant mass prescriptions following \cite{Fryeretal2012} and here we choose the delayed supernova (SNe) mechanism as the slow extreme of the convection-enhanced neutrino-driven SNe paradigm. We use standard momentum conserving fallback-scaled kicks (drawn from a Maxwellian distribution with a dispersion of 265.0~$\mathrm{kms}^{-1}$ from \citet{Hobbsetal2005}) for the neutron stars (NSs) and black holes (BHs) \citep{Belczynskietal2008}, except for the NSs and BHs that are produced by the electron-capture SNe (ECSNe), accretion-induced collapse (AIC) and merger-induced collapse (MIC) \citep{Podsiadlowskietal2004b,Ivanovaetal2008,GessnerJanka2018,Leungetal2020a} and that are subject to low velocity kicks (drawn from a Maxwellian distribution with a dispersion of 3.0~$\mathrm{kms}^{-1}$ from \citet{GessnerJanka2018}). The BHs receive natal spins following the \texttt{Geneva} models \citep{Banerjeeetal2020,Banerjee2021a}. The white dwarfs (WDs) receive natal kicks following \citet{Fellhaueretal2003} (drawn from a Maxwellian distribution with a dispersion of 2.0~$\mathrm{kms}^{-1}$ but, which is capped at 6.0~$\mathrm{kms}^{-1}$). We switch on the (pulsational) pair instability SNe following \citet{Belczynskietal2016}.

	\section{Results}
	\label{Section:Results}
	\subsection{Global dynamical evolution}
	\subsubsection{Structural parameter evolution}
	\label{Section:Structural parameter evolution}
	\begin{figure}
		\includegraphics[width=\columnwidth]{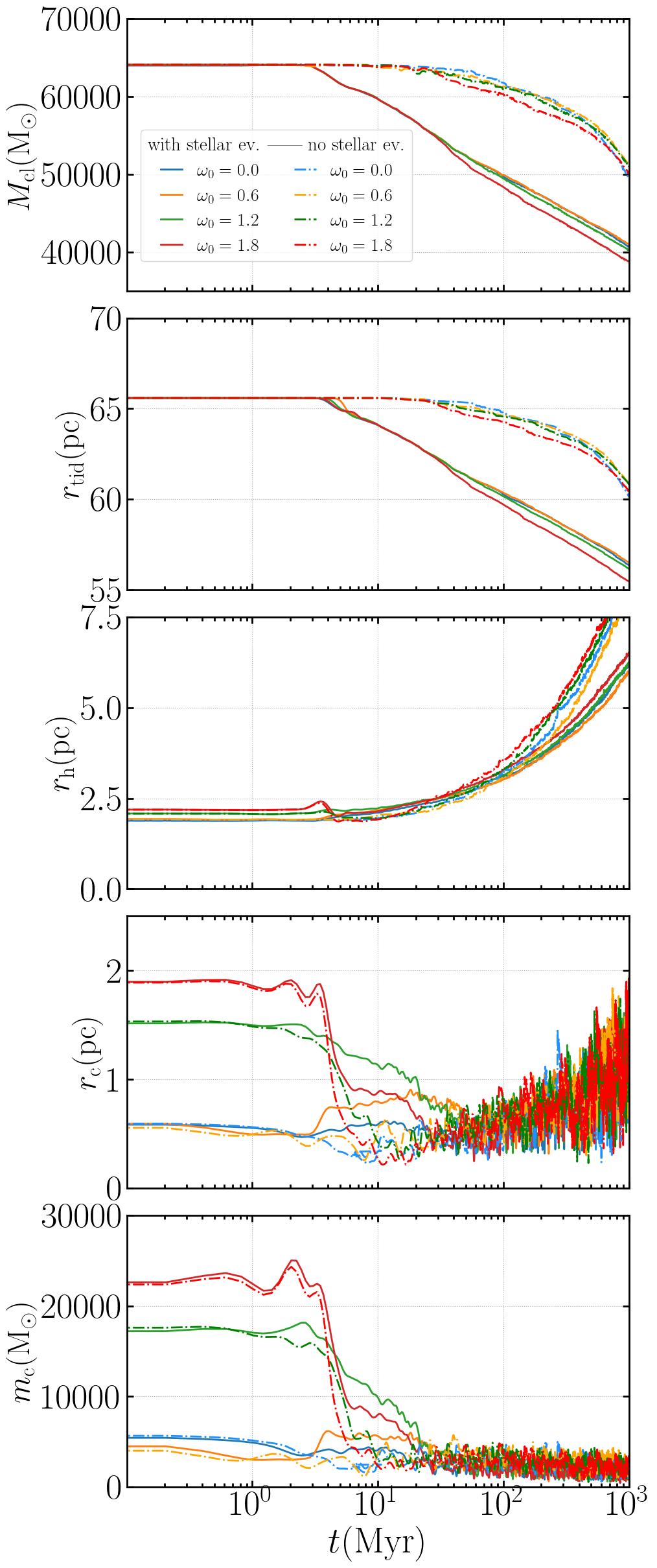}
		\caption{Plot showing the total cluster mass $M_{\mathrm{cl}}~~(\mathrm{M}_{\odot})$, the tidal radius $r_{\mathrm{t}}$~(pc), the half mass radius $r_{\mathrm{h}}$~(pc) and the mass of the core $m_{\mathrm{c}}~(\mathrm{M}_{\odot})$ and the radius of the core $r_{\mathrm{c}}$~(pc) in the four panels for all eight simulations with and without stellar evolution for $\omega_{0}=0.0, 0.6, 1.2, 1.8$, respectively. The time axis is plotted logarithmically to show the details of the much more rapid early cluster evolution. The models with stellar evolution (\texttt{SEV} models) are plotted as solid lines and the models without stellar evolution (\texttt{noSEV}) runs are plotted as dash-dotted lines.}
		\label{Global_properties_newtidal}
	\end{figure}
	We run each of the four initial models ($\omega_{0}=0.0, 0.6, 1.2, 1.8)$ with \textsc{Nbody6++GPU} once with stellar evolution switched on (\texttt{SEV} models) and once without (\texttt{noSEV} models). Hence we have eight distinct simulations to compare and contrast. We discuss in the following Figs.~\ref{Global_properties_newtidal} to \ref{BH_bar}, to get an overview on the global evolution of the simulated star clusters.\\
	Fig.~\ref{Global_properties_newtidal} shows the total cluster mass $M_{\mathrm{cl}}~(\mathrm{M}_{\odot})$, the tidal radius $r_{\mathrm{t}}$~(pc), the half mass radius $r_{\mathrm{h}}$~(pc), the mass of the core $m_{\mathrm{c}}~(\mathrm{M}_{\odot})$ and the radius of the core $r_{\mathrm{c}}$~(pc) in the four panels, respectively. In \textsc{Nbody6++GPU}, particles (single or binary stars) are removed from the star cluster once they have reached a distance that is twice the \textit{current} tidal radius far away from the density centre. They are called 'escapers' thereafter. The current tidal radius is then calculated using the \textit{current} cluster mass. Escapers do not contribute to the current cluster mass. They are also not taken into account when calculating any of the other structural parameters of the star clusters, such as $r_{\mathrm{h}}$ or $m_{\mathrm{c}}$. \\
	First, we look at the time evolution of $M_{\mathrm{cl}}$ and $r_{\mathrm{t}}$ for all eight models. While $M_{\mathrm{cl}}$ and $r_{\mathrm{t}}$ decrease significantly due to stellar evolution mass loss in the \texttt{SEV} models, the \texttt{noSEV} models can only suffer mass loss through escaping stars, either through strong dynamical encounters or series of weak encounters. It is therefore unsurprising that in the presence of the additional mass loss mechanism through stellar evolution, the tidal radii of the respective \texttt{SEV} models exhibit a much faster decrease. We also observe that the \texttt{noSEV} appear to approach the \texttt{SEV} counterpart models in their tidal radii in the indicating that the cluster evolution is faster in the long-term. We need simulations longer than 1~Gyr to make a more qualified statement on this.\\
	The half-mass radii $r_{\mathrm{h}}$ show an interesting evolution in time. While the evolution over the first couple of hundred Myrs is similar, the \texttt{noSEV} clearly diverge from the \texttt{SEV} models, which means the \texttt{noSEV} expand faster and more violently than the \texttt{SEV} models. This evolution is not mirrored by the core radius $r_{\mathrm{c}}$ evolution, which is similar in the longer term leading up to 1~Gyr. There is one striking difference though. All \texttt{noSEV} models collapse faster and exhibit a stronger core collapse than their counterparts with stellar evolution. However, the mass in the core evolves similarly meaning that the core mass $m_{\mathrm{c}}~(\mathrm{M}_{\odot})$ decreases faster and more strongly in all \texttt{noSEV} models. The evolution of the core radii and core masses are occur approximately synchronised, in all simulations. \\
	The time evolution of the Lagrangian radii $r_{\mathrm{Lagr}}$ or more precisely, the radii of mass shells containing a certain percentage of the \textit{current} total cluster mass (in this paper 1~\%, 5~\%, 10~\%, 30~\%, 50~\%, and 90~\% are shown), and the time evolution of the average stellar mass within these Lagrangian radii $M_{\mathrm{av}}$ are shown in Fig.~\ref{Lagrangian} for the all eight simulations. Each of the four columns represents a rotational parameter ($\omega_{0}$=0.0, 0.6, 1.2, 1.8) and every second row shows the \texttt{noSEV} models on a light grey background. It appears that the core-collapse phase of the star cluster \texttt{noSEV} models is more extreme, while the overall collapse also happens earlier. This observation is especially clear in the plots of $M_{\mathrm{av}}$ in the bottom two rows of Fig.~\ref{Lagrangian}, which shows a much faster mass segregation in the \texttt{noSEV} than in the \texttt{SEV} models. Moreover, the expansion of the outer-most Lagrangian radii happens significantly faster in the \texttt{noSEV} than in the \texttt{SEV} models, which adds further evidence for a faster evolution of the \texttt{noSEV} models. \\
	Overall, the discussion above can be related to the theorems described already in \cite{Henon1975} (see also \cite{BreenHeggie2013}). The evolution of the cluster system as a whole is governed by the energy flow through the half-mass radius $r_{\mathrm{h}}$ and it is independent of internal energy sources. The energy flow is approximately equal to $(GM_{\mathrm{cl}}^2/ r_{\mathrm{h}}) / t_{r_{\mathrm{h}}}$, where $t_{r_{\mathrm{h}}}$ is the half-mass relaxation time-scale and $M_{\mathrm{cl}}$ is the cluster mass, and this is equal to the energy generated at the centre of the cluster. In general, stellar evolution causes mass loss and results in an increase of $r_{\mathrm{h}}$. Additionally, the loss of mass by interaction and relaxation for very massive stars (without evolution) causes an increase in $r_{\mathrm{h}}$. Because in the case of no evolution we have more massive stars than in the case of evolution, the core collapses deeper and earlier. Mass loss through evolution slows down the collapse that then continues further. To stop the core collapse (no evolution), it is necessary to eject out some of the most massive binary systems and the most massive stars (as can be seen in the following figures). Then equilibrium occurs and both systems evolve similarly at the centre, generating similar energy. So if the mass of the system without stellar evolution is greater, then $r_{\mathrm{h}}$ must also be greater than in the case with stellar evolution. \\	
	Here, we also need to point out an important caveat: technically, as was also briefly outlined in Sect.~\ref{Section:Methods}, it is not entirely accurate to use $r_{\mathrm{c}}$, $r_{\mathrm{h}}$ and $r_{\mathrm{Lagr}}$ as measures for the global structure evolution of the rotating star cluster models that deviate too far from spherical symmetry. Instead of using Lagrangian mass shells, it would be better to sort the particles in terms of binding energy. This procedure would yield spheroids of equipotential surfaces. With these, we would then be able to calculate the respective radii along the principal axes of the spheroid, which is done below for the investigation of shape evolution of the star cluster models. \\
	As was outlined in Sect.~\ref{Section:Introduction}, bulk rotation leaves an imprint on the shape of a star cluster. In general, the flattening of a rotating mass distribution can be calculated by transforming the principal axes of a the moment of inertia tensor relative to the density centre of the mass distribution using different numbers of particles which are sorted by their binding energy \citep{TheisSpurzem1999}. Fig.~\ref{triaxility} shows the principal axis ratios of the intermediate to major axis ratio  $b/a$ and the minor to the major axis ratio $c/a$. Furthermore, following \citet{TheisSpurzem1999}, we define a triaxiality parameter of the system
	\begin{equation}
	\tau = \frac{b-c}{a-c},
	\end{equation}
	which is shown in the bottom two rows of Fig.~\ref{triaxility} (in this paper 10~\%, 30~\%, 50~\%, and 90~\% are shown).  As in Fig.~\ref{Lagrangian}, the \texttt{noSEV} models are plotted in a light-grey background. 
	We note that stochastic $N$-body noise disturbs the clean numbers. First, in the inner shells just the particle numbers are small. Second, our program does not have a fixed orientation for $a$, $b$ and $c$; the principal axes analysis always computes three principal axes and sorts them according to size. Therefore, stochastic noise always leads to $b/a$ and $c/a$ to be a bit smaller than unity, never greater. Stochastic noise in these quantities is also increased by the presence of massive stars, binaries, and fast evolving stellar masses (stellar evolution). For this reason, we have also refrained from plotting any shells below 50~\% in this paper. We would need much larger particle numbers than $1.1\times 10^5$ that we use in this work to have a more robust calculation that is less affected by these effects. Additionally, we note that the values of $\tau$ in Fig.~\ref{triaxility} are unreliable, because the definition of tau is not suitable for nearly spherical systems with $b\sim c$ and $a\sim c$.\\
	Overall, the impact of the stellar evolution in combination with tidal field mass loss from the cluster is significant. While the \texttt{SEV} models return from the maximum triaxiality ($b\neq a\neq c$) at minimum $c/a$ and $b/a$ to axisymmetry ($b=a$, but $c\neq a$ and $c\neq b$), the star clusters without stellar evolution activated do not exhibit this evolution. In fact, all \texttt{noSEV} models show the initial maximum triaxiality earlier and more pronounced than the \texttt{SEV} models and while they then shortly after are attempting to return to axisymmetric configurations, they then show no, one or two consecutive triaxial "collapses" ($\tau$ in bottom row of Fig.~\ref{triaxility}).
	\begin{figure*}
		\includegraphics[width=\textwidth]{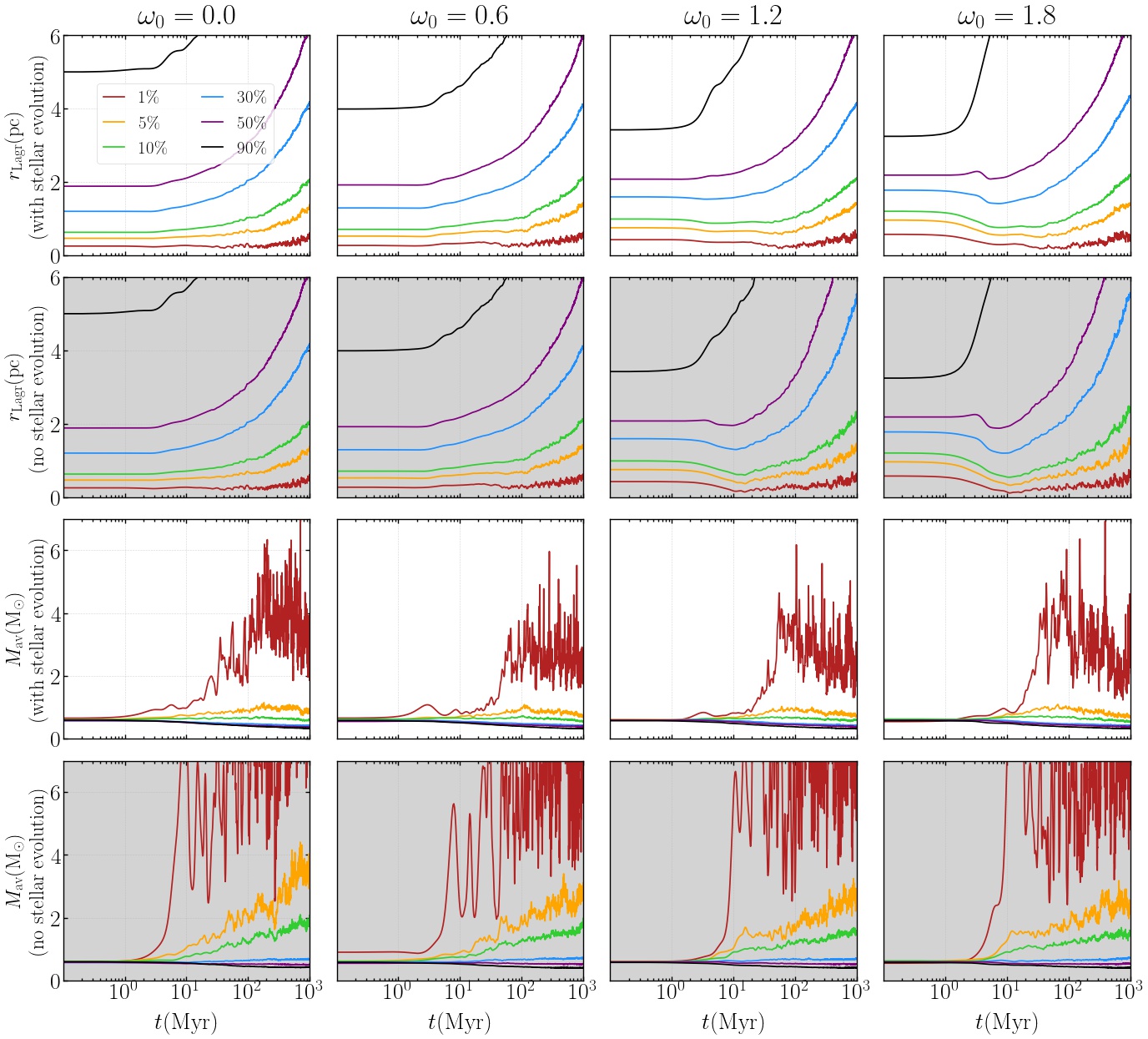}
		\caption{Plot showing the Lagrangian radii $r_{\mathrm{Lagr}}$~(pc) and the average mass $M_{\mathrm{av}}$~$(\mathrm{M}_{\odot})$ within shells that contain 1\%, 5\%, 10\%, 30\%, 50\%, and 90\% of the total cluster mass at the current simulation time step for up to 1~Gyr. The time axis is plotted logarithmically to show the details of the much more rapid early cluster evolution. Each column represents one rotational parameter $\omega_{0}$ of the rotating King model in ascending order from left to right ($\omega_{0}$=0.0, 0.6, 1.2, 1.8). The results from the runs \textit{with} stellar evolution switched on (\texttt{SEV} models) are plotted on a white background, while the results from the simulations \textit{without} stellar evolution (\texttt{noSEV} models) are highlighted in light grey.}
		\label{Lagrangian}
	\end{figure*}
	\begin{figure*}
		\includegraphics[width=\textwidth]{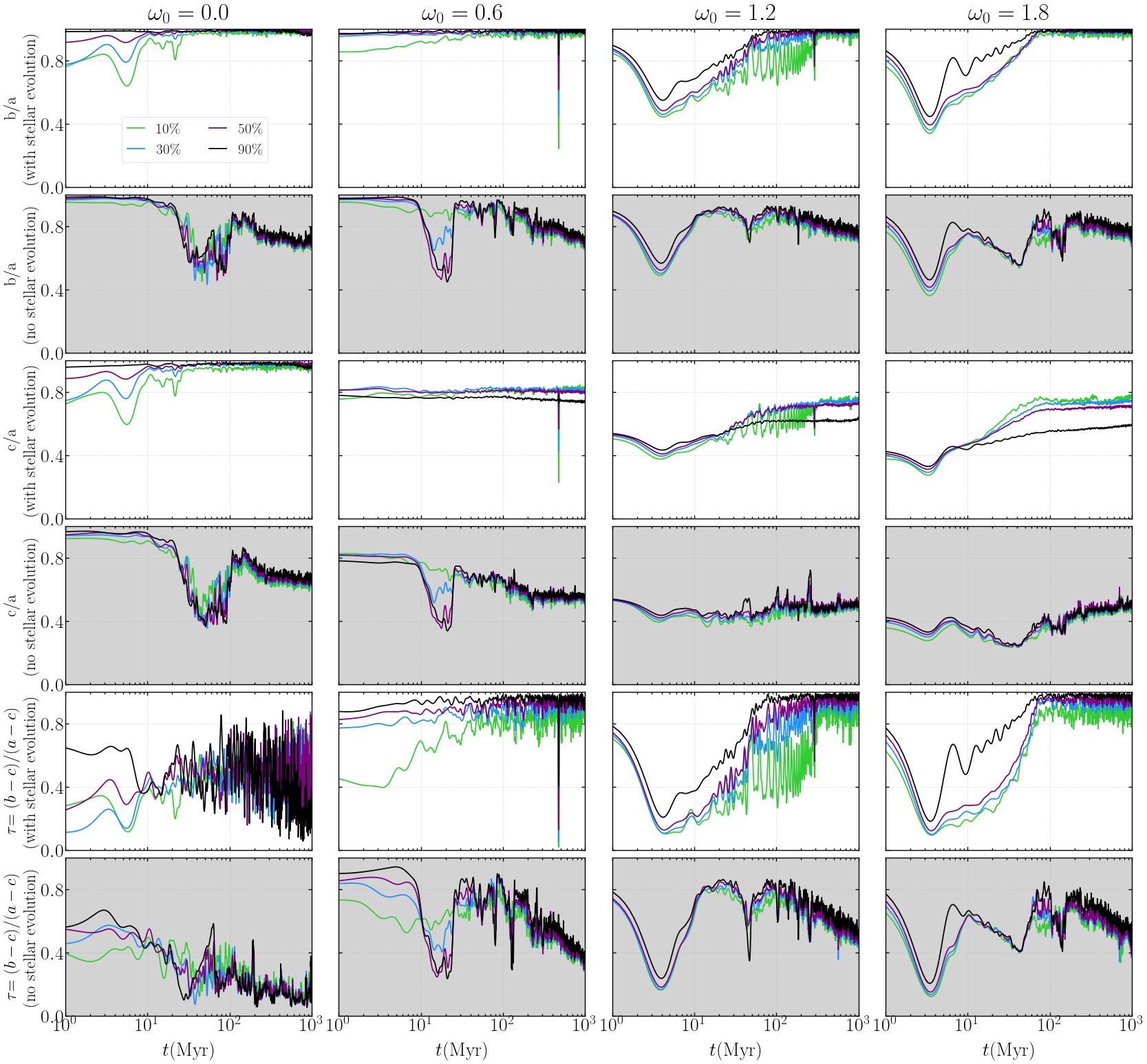}
		\caption{Plot showing the ratios of the principal axis of the moment of inertia tensor, $b/a$ and $c/a$, as well as the triaxiality parameter $\tau = (b-c)/(a-c)$ within shells that contain 10\%, 30\%, 50\%, and 90\% of the total particle energy at the current simulation time step for up to 1~Gyr. The time axis is plotted logarithmically to show the details of the much more rapid early cluster evolution. Each column represents one rotational parameter $\omega_{0}$ in ascending order from left to right ($\omega_{0}$=0.0, 0.6, 1.2, 1.8). The results from the runs \textit{with} stellar evolution switched on (\texttt{SEV} models) are plotted on a white background, while the results from the simulations \textit{without} stellar evolution (\texttt{noSEV} models) are highlighted in light grey.}
		\label{triaxility}
	\end{figure*}
	\begin{figure*}
		\includegraphics[width=\textwidth]{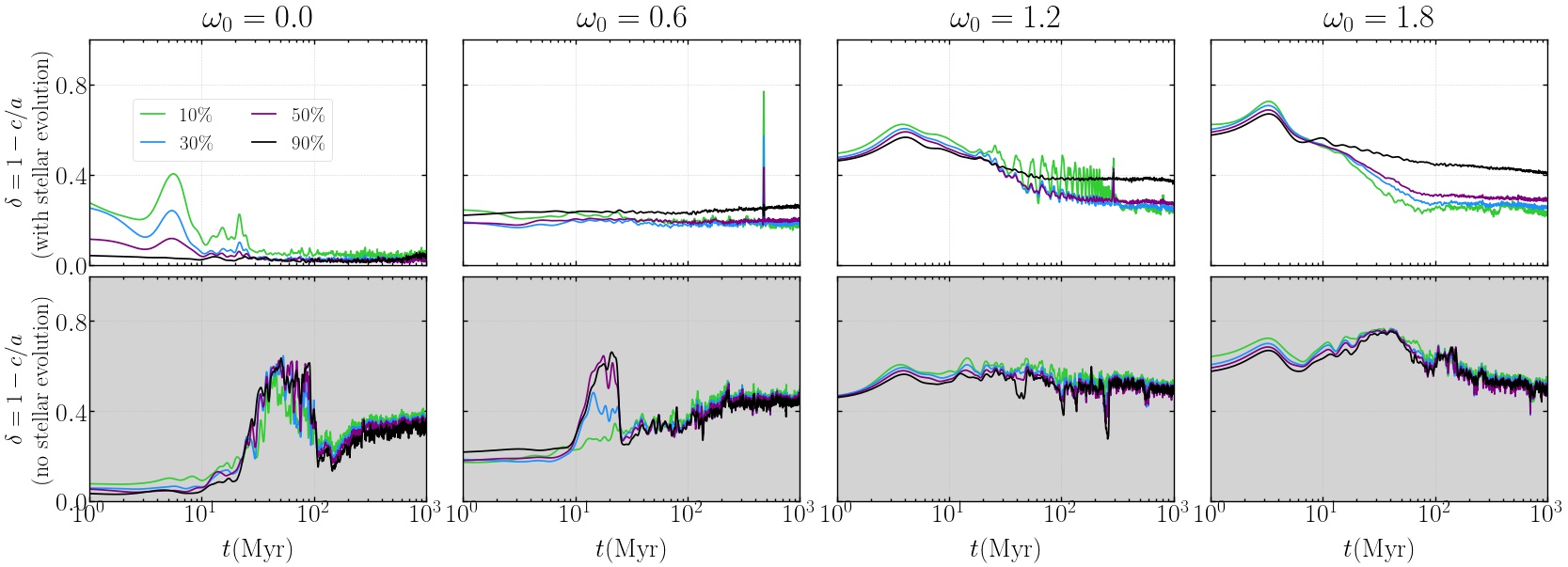}
		\caption{Plot showing the deviation from sphericity of the star cluster models, $\delta = 1-c/a$, within shells that contain 10\%, 30\%, 50\%, and 90\% of the total particle energy at the current simulation time step for up to 1~Gyr. The time axis is plotted logarithmically to show the details of the much more rapid early cluster evolution. Each column represents one rotational parameter $\omega_{0}$ in ascending order from left to right ($\omega_{0}$=0.0, 0.6, 1.2, 1.8). The results from the runs \textit{with} stellar evolution switched on (\texttt{SEV} models) are plotted on a white background, while the results from the simulations \textit{without} stellar evolution (\texttt{noSEV} models) are highlighted in light grey.}
		\label{Global_sphericity}
	\end{figure*}
	Furthermore, it is noteworthy that all shells from $10~\%$ to $90~\%$ are much more similar in their structure evolution for the \texttt{noSEV} compared with their counterparts in the \texttt{SEV} models, where there is more divergence between individual spheroidal shells. This is possibly related to the tidal field mass loss, meaning that if the tidal radius was (much) larger, the \texttt{noSEV} would show a similar evolution compared with the \texttt{SEV} models. \\
	From Fig.~\ref{triaxility} and Fig.~\ref{Lagrangian} we can deduce the following cluster evolution qualitatively. 
	Let us first look at the rotating clusters ($\omega_0 > 0.0$).
    First, there is a strong core collapse, which can be identified by the first
    maximum of the average mass in Fig.~\ref{Lagrangian}; it is earlier for \texttt{noSEV} runs, because they keep high stellar masses and thus experience fast mass segregation. For \texttt{SEV} runs
    heavy masses evolve fast, have strong mass loss, so collapse by mass segregation
    is slower. It is interesting to note that approximately at the first core
    collapse there is a {\em minimum} value of triaxiality $\tau$ and $\delta$ shown in Fig.~\ref{Global_sphericity} ($\delta = 1-c/a$, a measure of flattening between the major and minor axes ($a$ and $c$), it is 0 for spherical systems, and one for disky systems, see also \citet{TheisSpurzem1999}). That is followed a couple of Myrs later by a strong maximum in both $\tau$ and $\delta$. We interpret this as follows: during collapse at high density the relaxation time is short, the system is developing towards sphericity and isotropy. Afterwards a radial orbit instability (ROI) is developing which produces the maximum of $\tau$ and $\delta$; the ROI is stronger for faster rotation, because we have less energy in the tangential unordered motion (tangential velocity dispersion becomes smaller compared to rotational velocity). Here, we did not examine in more detail the onset of ROI, the interested reader is referred to \citet{TheisSpurzem1999} and earlier references therein. \\
     For the non-rotating system there is also a core-collapse by mass segregation, faster in the \texttt{noSEV} case than with \texttt{SEV}; opposite to expectation the system develops some non-sphericity, in the case of \texttt{noSEV}$\omega_00.0$.\\
    Second, we find a phase of restoration of axisymmetry for the SEV models. The outermost shells exhibit oscillations in shape that are dampened over time, and the system returns to a stationary, flattened, axisymmetric state ($\tau\sim 1$, $\delta>0$). It is interesting to note that the \texttt{noSEV} model does not return to axisymmetry, on the contrary it keeps some triaxiality during the last few 100 Myrs of our simulation. The effect is more pronounced for the rotating systems, but as discussed before, the values of $\tau$ for non-rotating models should be taken with care. Why this is the case is currently unclear. Possible speculative explanations are ongoing repeated ROI due to central core oscillations supported by the heavy masses, or interactions of the external tidal field, removing angular momentum  (see Sect.~\ref{Section:Angular momentum evolution}).
	
	\subsubsection{Angular momentum evolution}
	\label{Section:Angular momentum evolution}
	\begin{figure*}
		\includegraphics[width=\textwidth]{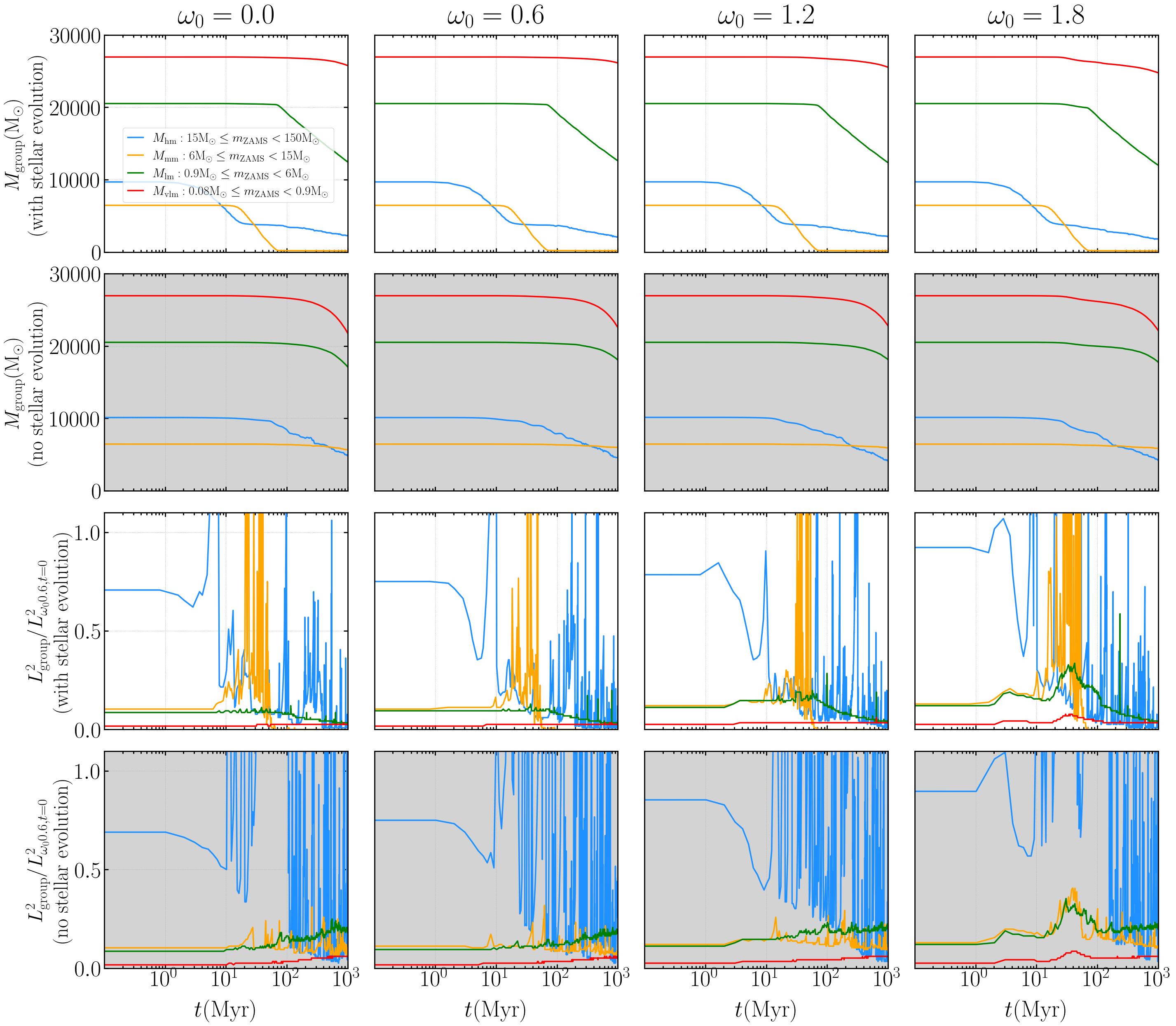}
		\caption{Plot showing the total mass of the four mass groups ($M_{\mathrm{vlm}},M_{\mathrm{lm}},M_{\mathrm{mm}},M_{\mathrm{hm}}$) in the top two rows and the square of the total angular momentum for these groups divided by the square of the total angular momentum of the $\omega_{0}0.6$ model(s) at $t=0$, $L_{\mathrm{group}}^2/L^2_{\omega_{0}0.6,t=0}$, at the current simulation time step for up to 1~Gyr. The time axis is plotted logarithmically to show the details of the much more rapid early cluster evolution. Each column represents one rotational parameter $\omega_{0}$ in ascending order from left to right ($\omega_{0}$=0.0, 0.6, 1.2, 1.8). The results from the runs \textit{with} stellar evolution switched on (\texttt{SEV} models) are plotted on a white background, while the results from the simulations \textit{without} stellar evolution (\texttt{noSEV} models) are highlighted in light grey.}
		\label{Angular_momentum_comparison}
	\end{figure*}
	We want to explore how the angular momentum is transported within the star cluster simulations and if and how this depends on the stellar evolution and initial bulk rotation strength. For this purpose, we divide the complete ZAMS particle set into four distinct mass groups (very low mass (vlm), low mass (lm), medium mass (mm) and high mass (hm)):
	\begin{align*} 
	M_{\mathrm{vlm}} &:  0.08~\mathrm{M}_\odot \leq m_{\mathrm{ZAMS}} < 0.9~\mathrm{M}_\odot \\ 
	M_{\mathrm{lm}} &:  0.9~\mathrm{M}_\odot \leq m_{\mathrm{ZAMS}} < 6~\mathrm{M}_\odot \\ 
	M_{\mathrm{mm}} &:  6~\mathrm{M}_\odot \leq m_{\mathrm{ZAMS}} < 15~\mathrm{M}_\odot \\ 
	M_{\mathrm{hm}} &: 15~\mathrm{M}_\odot \leq m_{\mathrm{ZAMS}} < 150~\mathrm{M}_\odot,
	\end{align*}
	where $m_{\mathrm{ZAMS}}$ is the ZAMS stellar mass of a single star (this also means that a primordial binary star could have binary members that are in two different mass groups). The mass groups are chosen such that the stars from $M_{\mathrm{hm}}$ become BHs, the stars from $M_{\mathrm{mm}}$ become NSs, the stars from $M_{\mathrm{lm}}$ become WDs and the stars from $M_{\mathrm{vlm}}$ remain as MSs for the simulation time, approximately. We can then follow the particles that originate from these mass groups through the full cluster evolution and compute their angular momentum across the full evolution. As a result, we are in a position to plot the time evolution of, for example, the square of the total angular momentum $L^2$ for each of the four mass groups and compare them to follow the angular momentum transfer. In Cartesian coordinates, $L^2$ for an individual star is simply given as quadratic sum of three components
	\begin{align}
	L^2_{x} &= (yp_{\mathrm{z}}-zp_{\mathrm{y}})^2  ,\\
	L^2_{y} &= (zp_{\mathrm{x}}-xp_{\mathrm{z}})^2  ,\\
	L^2_{z} &= (xp_{\mathrm{y}}-yp_{\mathrm{x}})^2  ,
	\end{align}
	which can then be done for all stars in each individual mass group. The sum of $L^2$ of all individual stars then gives the $L_{\mathrm{group}}^2$, the total sum of the square of the angular momentum. \\
	All $L_{\mathrm{group}}^2$ are divided by $L^2_{\omega_{0}0.6,t=0}$, which is the square of the total angular momentum of the $\omega_{0}0.6$ model(s) at $t=0$ (the sum of all $L_{\mathrm{group}}^2$ for the $\omega_{0}0.6$ models divded by $L^2_{\omega_{0}0.6,t=0}$ is one). We do this so that the models can be compared with each other more easily. $L_{\mathrm{group}}^2/L^2_{\omega_{0}0.6,t=0}$ is shown in Fig.~\ref{Angular_momentum_comparison} for all models. $M_{\mathrm{group}}$, which is the mass of all the stars (and compact objects) in the four groups as a function of time, is also shown in Fig.~\ref{Angular_momentum_comparison}. First of all, we see that the total mass in each mass group evolves similarly at least initially across the \texttt{SEV} and across the \texttt{noSEV} models until stellar evolution and associated mass loss take over. With increasing initial bulk rotation, the mass loss from particularly the mass group of very low mass stars, $M_{\mathrm{vlm}}$, is enhanced. This mass loss is assisted due to mass segregation and therefore, it is unsurprising that $M_{\mathrm{vlm}}$ is especially affected by this, because the member stars migrate to the cluster halo over time. The \texttt{noSEV} models lose mass only via tidal field mass loss or due to strong few-body encounters in the central high density region, which kick out stars and
    lift them up to escape energies. They also lose more mass by escaping stars than the \texttt{SEV} models (see Fig.~\ref{escaper_properties} in Sect.~\ref{Section:Escaper stars}). Due to stellar evolution, the \texttt{SEV} models lose mass in all mass groups much earlier during the simulation. It is especially striking in the medium mass $M_{\mathrm{mm}}$ and high mass $M_{\mathrm{hm}}$ groups, which predominantly produce NSs and BHs, respectively. \\
	We now discuss the evolution of the angular momentum of the mass groups with the quantity $L_{\mathrm{group}}^2/L^2_{\omega_{0}0.6,t=0}$, which reveals an important result that is particularly clear for increasing initial bulk rotation. From Fig.~\ref{escaper_properties} we can qualitatively conclude the angular momentum loss and exchange - the angular momentum lost by the heavy mass group goes into cluster mass loss in the non- or slowly rotating case, only little is transferred to the light mass groups. The relative importance can be estimated from Fig.~\ref{escaper_properties}, which compares the mass loss for \texttt{noSEV} and \texttt{SEV} models. The interesting finding here is, however, that for the highly rotating systems a larger fraction of the heavy mass angular momentum is transferred to the light mass groups (but finally they also lose angular momentum due to general cluster mass loss). This is a signature of gravogyro catastrophe.
 \\
	The spikes in the $L_{\mathrm{group}}^2/L^2_{\omega_{0}0.6,t=0}$ curves are due to escaping stars or compact objects, which gain large amounts of angular momentum and then escape the cluster. It is important to keep in mind here that compact objects receive natal kicks in our simulations. Therefore, the number of these spikes is much higher in the \texttt{SEV} models (see in particular for the $M_{\mathrm{mm}}$), because in the \texttt{noSEV} models, the stars can only escape through dynamical interactions. We can particularly see this in the evolution of the $L_{\mathrm{mm}}^2/L^2_{\omega_{0}0.6,t=0}$ and comparing it between the \texttt{noSEV} and \texttt{SEV} models. Remember that the objects from this group produce mostly NSs that receive very large natal kicks (several hundreds of $\mathrm{kms}^{-1}$). We see that in the intermediate to long term of our simulations, the angular momentum loss from the \texttt{SEV} is much larger than that from the \texttt{noSEV} models, which becomes especially clear for the models with very large initial bulk rotation. While the \texttt{noSEV} models have a roughly constant angular momentum evolution above 100~Myr for the $M_{\mathrm{vlm}}$, $M_{\mathrm{lm}}$ and $M_{\mathrm{mm}}$ mass groups, the \texttt{SEV} models show a clear decrease of angular momentum in all four mass groups. This effect is achieved through angular momentum loss through escaping stars and mass loss due to stellar evolution.\\
	We also see for the \texttt{noSEV} models that when comparing Fig.~\ref{Angular_momentum_comparison} with Fig. \ref{Lagrangian} and Fig. \ref{triaxility}, it becomes clearer that the \texttt{noSEV} models are unstable in their global evolution for all four runs ($\omega_{0}$= 0.0, 0.6, 1.2, 1.8). By increasing the initial tidal radius in future simulations, this might be a very different situation. \\
	Lastly, Fig.~\ref{Angular_momentum_comparison} reveals another important result. In the following discussion we focus on the $M_{\mathrm{mm}}$ mass group in the \texttt{noSEV} models. We can see that this group consistently has an almost constant mass ($M_{\mathrm{group}}$; with very small fluctuations). It appears that stars from this mass group are not ejected from the cluster. Furthermore, we see from $L_{\mathrm{group}}^2/L^2_{\omega_{0}0.6,t=0}$ for this mass group that its angular momentum effectively approaches zero after a couple of Myrs. This process can imply that the $M_{\mathrm{mm}}$ objects replace the depleting numbers of $M_{\mathrm{hm}}$ objects in the cluster centre in the mid- to long-term cluster evolution (see \cite{Contentaetal2015} for the formation of a NS subsystem in the cluster centre). \\
	Here, we also need to add an important caveat: the angular momenta are computed relative to the cluster density centre. However, since with have a tidal field the whole cluster experiences a (small) recoil every time a particle escapes by nature of momentum conservation. Therefore, the cluster density centre might move relative to the cluster centre of mass, which would have a (small) effect on the computation of the angular momentum. 
	
	\subsubsection{Bar and disk formation of heavy mass objects}
	\begin{figure*}
		\includegraphics[width=\textwidth]{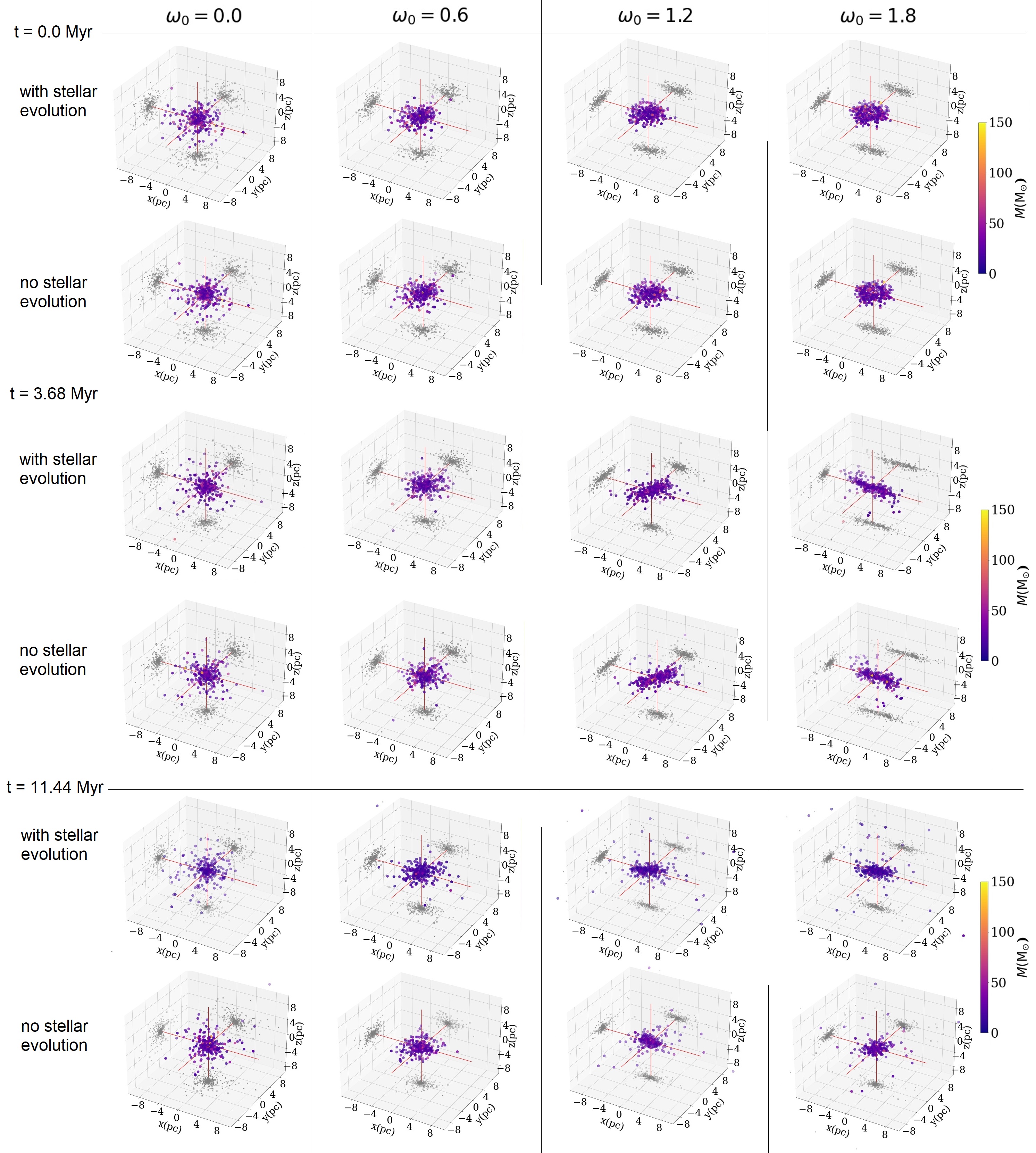}
		\caption{3-D scatter plot showing the spatial distribution of the $M_{\mathrm{hm}}$ mass group in all eight simulations at 0.0~Myr, 3.68~Myr, and 11.44~Myr from the top to bottom in three separate rows with two sub-rows each; the top sub-row are always the models \textit{with} stellar evolution (\texttt{SEV} models) and the bottom sub-row are always the models \textit{without} stellar evolution (\texttt{noSEV} models). There are four columns and each one represents a rotational parameter $\omega_0$ in ascending order of rotation from left to right. The stars and compact objects are color-coded by their mass between 0.0~$\mathrm{M}_{\odot}$ and 150.0~$\mathrm{M}_{\odot}$. The stars and BHs are also projected onto the three dimensional axes, which can be seen from the light-grey dots. We can clearly see the bar formation of the BHs and their progenitor stars in
			at t=3.68~Myr and the spatial reconfiguration of the $M_{\mathrm{hm}}$ objects to axisymmetric structures.}
		\label{BH_bar}
	\end{figure*}
	Here, we explore the spatial evolution of the high mass group $M_{\mathrm{hm}}$. We want to know what happens to the shape of the distribution of these objects and how it is affected by initial bulk rotation and stellar evolution. In the \texttt{SEV} models this corresponds to the shape of the distribution of the BHs and their progenitor stars. Fig.~\ref{BH_bar} shows the 3-D spatial distribution of the stars and compact objects from $M_{\mathrm{hm}}$ at 0.0~Myr, 3.68~Myr, and 11.44~Myr from top to bottom, respectively. This time is approximately the time of maximum triaxiality for the \texttt{SEV}$\omega_01.8$ model (meaning approximately the simulation snapshot that is closest to maximum triaxiality). The bar formation of the BHs and their progenitor stars is clear in the \texttt{SEV}$\omega_01.2$ and \texttt{SEV}$\omega_01.8$ models (see also \cite{Hongetal2013} for more on bar formation). Their \texttt{noSEV} model counterparts, \texttt{noSEV}$\omega_01.2$ and \texttt{noSEV}$\omega_01.8$, also show the formation of a bar. It seems to be similar in spatial distribution, however, we know already from Fig.~\ref{triaxility} that the \texttt{noSEV} models do in fact yield slightly more maximally triaxial configurations. This overall process has also been referred to \textit{anisotropic mass segregation} in \citet{Szolgyenetal2021,PanamarevKocsis2022}. The \texttt{noSEV}$\omega_01.2$ and \texttt{noSEV}$\omega_01.8$ also attempt to return to axisymmetric configurations at 11.44~Myr. However, they seem to be slightly more concentrated than the \texttt{SEV} counterparts. We can infer on this from Fig.~\ref{Lagrangian}. This effect is also due stellar evolution mass loss, which is in turn related to the natal kicks that the BHs experience. Therefore, it is natural that you can see larger spatial scattering in the distributions regardless of $\omega_0$ compared to their \texttt{noSEV} model counterparts.  \\
	In summary, the initially rotating axisymmetric distribution of the $M_{\mathrm{hm}}$ objects becomes a bar that rotates around the z-axis and evolves toward a disc configuration over time (at least for the \texttt{SEV} models, see also Fig.~\ref{triaxility}). This is strictly not the case for $M_{\mathrm{hm}}$ objects in the non-rotating ($\omega_0=0.0$) models. Here, the \texttt{SEV} and \texttt{noSEV} models stay spherical at least for the first 11.44~Myr of the simulations. However, we know from Fig.~\ref{triaxility} that also the \texttt{noSEV}$\omega_00.0$ and \texttt{noSEV}$\omega_00.6$ deviate from spherical symmetry over time. This deviation in the respective \texttt{noSEV} models is due to enhanced tidal field mass loss and tidal tails in the cluster (see discussion in Sect.~\ref{Section:Structural parameter evolution} and Fig.~\ref{Global_sphericity}). \\ 
	Young open clusters would be an ideal target for observations and further simulations to test this experimental result. In \citet{Pangetal2022}, elongated shapes of young clusters of \textit{filamentary-type} might still carry the signal of a bar structure induced by rotation. However, the dynamical bar structure may blend with the inherent filamentary structure. We need to differentiate them carefully via kinematic data.
	
	\subsection{Escaper stars}
	\label{Section:Escaper stars}
	\begin{figure*}
		\includegraphics[width=\textwidth]{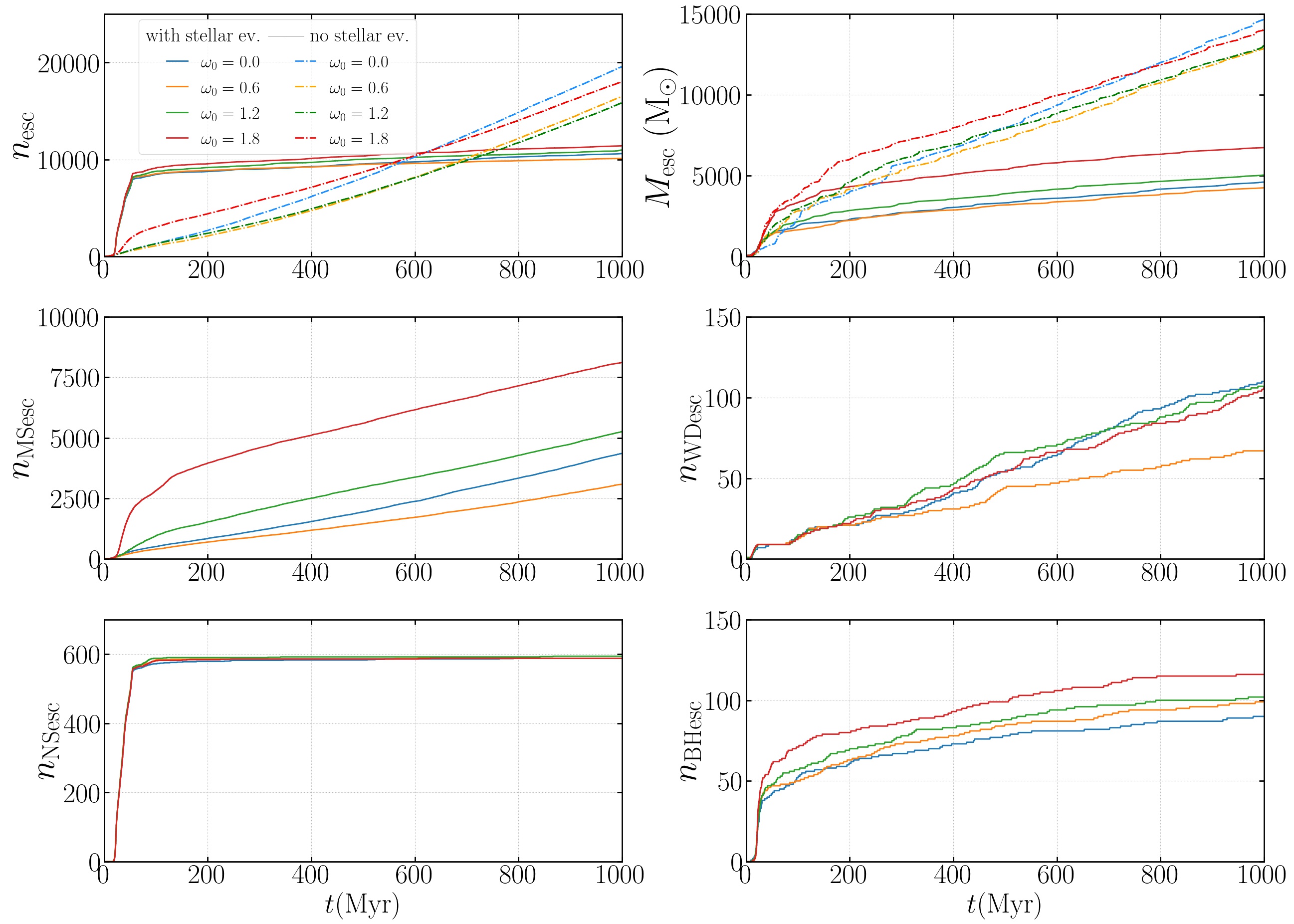}
		\caption{Plot showing over 1~Gyr the number of escapers, $n_{\mathrm{esc}}$, the total mass evolution of the escapers, $M_{\mathrm{esc}}$~($\mathrm{M}_{\odot}$), the number of escaping MS stars, $n_{\mathrm{MSesc}}$, the number of escaping WDs, $n_{\mathrm{WDesc}}$, the number of escaping NSs, $n_{\mathrm{NSesc}}$, and the number of escaping BHs, $n_{\mathrm{BHesc}}$, respectively. The latter four are naturally only shown for the \texttt{SEV} models.}
		\label{escaper_properties}
	\end{figure*}
\begin{figure}
		\includegraphics[width=\columnwidth]{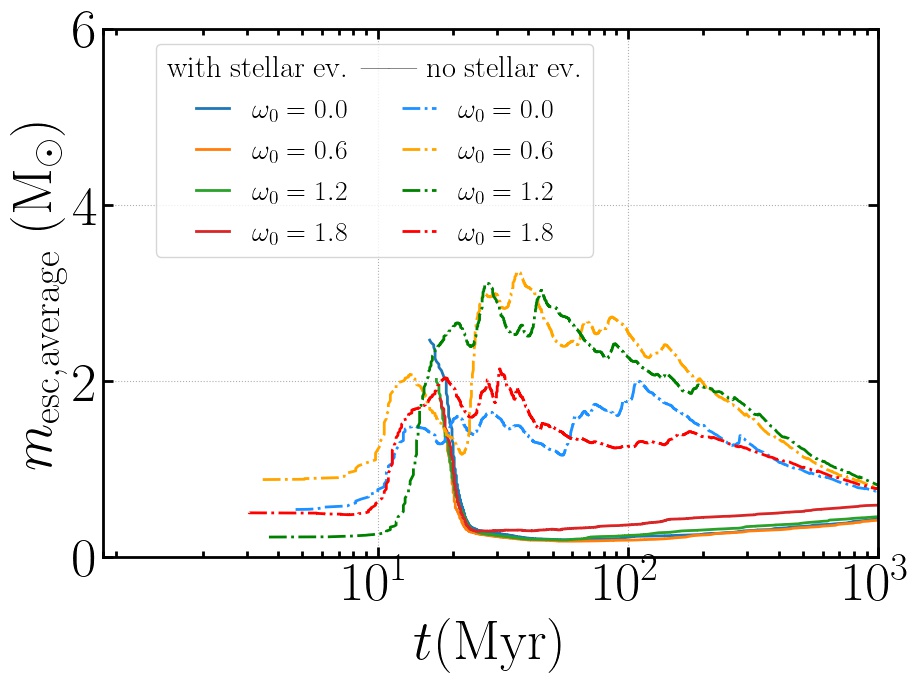}
		\caption{Plot showing the average mass of the escapers, $m_{\mathrm{esc,average}}$, over 1~Gyr of cluster evolution.}
		\label{average_mass_escapers}
	\end{figure}

	The escapers from the simulations reveal more important information and are shown in Fig.~\ref{escaper_properties}. In the following, we can study the temporal evolution of the number of escapers, $n_{\mathrm{esc}}$. The \texttt{SEV} models initially lose  more stars and compact objects than the \texttt{noSEV} models, but the \texttt{noSEV} models start losing stars significantly earlier, which is more apparent in the semi-logarithmic scaling in Fig.~\ref{average_mass_escapers}, which is due to the faster evolution of the \texttt{noSEV} models (see also Sect.~\ref{Section:Structural parameter evolution}). The initially strong increase in the number of escapers is due to the large cluster mass reduction (potential) and increase of a number of stars called potential escapers. However, depending on the initial rotation, the \texttt{noSEV} models produce more escapers after a couple of Myr of simulation time. The runs with larger initial bulk rotation lose more stars, which is also the case initially for the \texttt{noSEV} models. Here, the number of escapers of the runs without any rotation surpass the most strongly rotating run at about 600~Myr. We see a constant and almost linear rise of escaper numbers for the \texttt{noSEV} compared with the much flatter increase in escapers for the \texttt{SEV} models. We therefore confirm that the tidal field mass loss is much stronger for the \texttt{noSEV} models in the long-term, which can also be inferred from the time evolution of the total mass of the escapers, $M_{\mathrm{esc}}$. The overall mass is larger and increases much faster in the \texttt{noSEV} than in the \texttt{SEV} models. \\
	Interestingly, while the escaper numbers for the \texttt{SEV} models are very similar, the total mass loss is much larger for the \texttt{SEV}$\omega_01.8$ models than for the \texttt{SEV} models that rotate less strongly initially. These numbers can be attributed mostly to the much larger number of initially escaping MS stars, $n_{\mathrm{MSesc}}$, which is also shown in Fig.~\ref{escaper_properties}. The lower mass MS stars are driven onto large orbits around the density centre of the star cluster by having the angular momentum transported to them through the gravogyro catastrophe. We can also see this effect from Fig.~\ref{Angular_momentum_comparison}, which is discussed in Sect.~\ref{Section:Angular momentum evolution}. Interestingly, the \texttt{SEV}$\omega_00.6$ retains many more MS stars than the \texttt{SEV}$\omega_00.0$ model. This discrepancy is also mirrored by the number of escaping WDs, $n_{\mathrm{WDesc}}$. For the other runs, \texttt{SEV}$\omega_01.2$ and \texttt{SEV}$\omega_01.8$, these are approximately similar over 1~Gyr. The number of escaping NS, $n_{\mathrm{NSesc}}$, are practically identical. The reason for this is that the NSs that escape suffer from very large natal kicks and only those that form via ECSNe, AIC or MIC are retained in the cluster. Since the IMF is the same for all models, it is unsurprising that similar numbers are retained. This is not the case for the BHs. The plot for $n_{\mathrm{BHesc}}$ reveals that the \texttt{SEV}$\omega_01.8$ models lose the largest number of BHs by a considerable margin. It might be suspected that $n_{\mathrm{BHesc}}$ should be similar for all models just like the evolution of $n_{\mathrm{NSesc}}$. However, the double-core collapse hump in combination with the fallback-dependent scaling of the natal kicks produces a larger diversity (see also \cite{Fryeretal2012,Kamlahetal2021}). \\
	Fig.~\ref{average_mass_escapers} shows the average mass of the escapers $m_{\mathrm{esc,average}}$ for the \texttt{SEV} and \texttt{noSEV} models. Apart from the fact that stars escape the \texttt{noSEV} models earlier as was discussed above, $m_{\mathrm{esc,average}}$ is much larger in the \texttt{noSEV} than in the \texttt{SEV} models. We define $m_{\mathrm{esc,average}}$ as $M_{\mathrm{esc}}$ divided by $n_{\mathrm{esc}}$ at a specific point in time. Recall, that we use a IMF following \citet{Kroupa2001} between ($0.08-150$)~$\text M_{\odot}$ (see Tab.~\ref{Initial_conditions}). Our IMF produces an average ZAMS for our cluster of around 0.58~$\mathrm{M_{\odot}}$. We see that the stars that escape the \texttt{noSEV} models are on average much more massive than the average star in the cluster. Due to the convective angular momentum transport, which happens extremely quickly and which is more dominant for increasing rotation (already after 0.1 Myr, see Fig.~\ref{Angular_momentum_comparison}), many (very) low mass, medium mass stars are removed along with high mass stars in the \texttt{noSEV}$\omega_01.8$ model. This observation is mirrored in Fig.~\ref{escaper_properties}, where many more stars are removed for the \texttt{noSEV}$\omega_01.8$ models initially than the other \texttt{noSEV} models. This effect brings down the average mass of the escapers. However, the \texttt{noSEV}$\omega_00.6$ and the \texttt{noSEV}$\omega_01.2$ produce remarkably similar evolution of $m_{\mathrm{esc,average}}$. Averaging over more simulations would produce more reliable results.
	
	\subsection{Binary stars}
	\label{Section:Binary stars}
	\begin{figure*}
		\includegraphics[width=\textwidth]{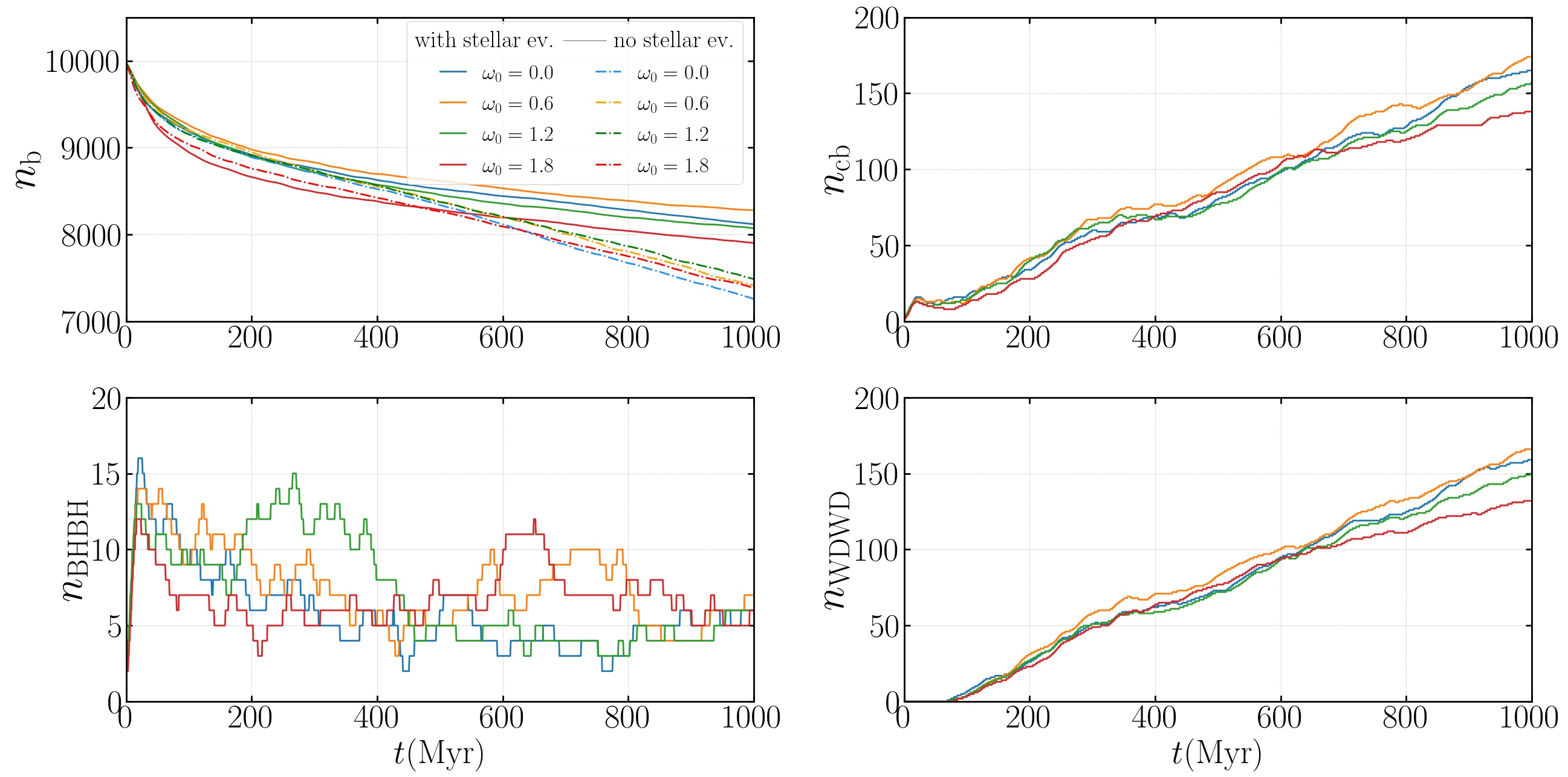}
		\caption{Plots showing the number of binary stars $n_{\mathrm{b}}$, the number of compact binary stars $n_{\mathrm{cb}}$, the number of binary black holes (BHBH) $n_{\mathrm{BHBH}}$, the number of binary white dwarfs (WDWD) $n_{\mathrm{WDWD}}$. For the plot of $n_{\mathrm{b}}$ both the \texttt{SEV} (solid lines) and the \texttt{noSEV} models (dash-dotted lines) are shown.}
		\label{Binary_abundances}
	\end{figure*}
	
	The temporal evolution of the number of binaries retained in the star clusters (both dynamical and primordial) can serve as a qualitative indicator for the number of dynamical interactions. Fig.~\ref{Binary_abundances} shows this number of binary stars $n_{\mathrm{b}}$ for all eight simulations. We first concentrate on the \texttt{SEV} models only. The \texttt{SEV}$\omega_01.8$ have considerably lower numbers of binaries at 1~Gyr than the other models, which can mostly be attributed to escaping or disrupted binaries (by stellar evolution or dynamical encounters) in the very early simulations. The other simulations show a similar evolution of $n_{\mathrm{b}}$ with the notable exception that $n_{\mathrm{b}}$ for \texttt{SEV}$\omega_00.6$ is larger than any of the other consistently over 1~Gyr. Now, comparing this with the evolution of $n_{\mathrm{b}}$ in the \texttt{noSEV} models, we find a different evolution. Here, the number of binaries show a lower scattering at 1~Gyr. Additionally, \texttt{noSEV}$\omega_01.8$ appears to produce an intermediate number of retained binary stars and the \texttt{noSEV}$\omega_00.0$ simulation produces the lowest numbers. To achieve greater clarity on this issue, we would need many simulations with different random realisations and look at the simulation ensemble average of the number of binaries for the different $\omega_0$ values. We would then be in a position if this is not a random effect or if there is some systematic evolution occurring. \\
	In the following discussion, we only consider the \texttt{SEV} models. The number of compact binaries, $n_{\mathrm{cb}}$, reveals that the \texttt{SEV}$\omega_01.8$ produce the lowest numbers of compact binaries retained in the cluster and the models with \texttt{SEV}$\omega_00.6$ retain the largest numbers of compact binaries, thereby mirroring the overall number of binaries retained in the cluster. $n_{\mathrm{cb}}$ consists practically only of BHBH and WDWD binaries in our simulations, which is also why only the number of BHBH binaries, $n_{\mathrm{BHBH}}$, and the number of WDWD binaries, $n_{\mathrm{WDWD}}$, are shown in Fig.~\ref{Binary_abundances}. Interestingly, there is a clear increase in the evolution of $n_{\mathrm{BHBH}}$ for the \texttt{SEV}$\omega_01.2$ model. This is important because it could indicate that IMBH formation might be preferential at this initial bulk rotation (note that the maximum of this increase is already much later than the dissolution of the bar structure and occurs when the clusters are axisymmetric again, see also Fig.~\ref{triaxility} and Fig.~\ref{BH_bar}). But it could also just be statistical fluctuation (compare this also to the smaller increases for the \texttt{SEV}$\omega_00.6$ and \texttt{SEV}$\omega_01.8$ models that occur later on). The number of BHs and BHBHs are both too low in our simulations to make a quantitative assessment on this. At 1~Gyr all simulations appear to converge at 5 or 6 BHBH binaries retained in the simulations. Our hypothesis here could be supported by the study of Brownian motion of BHs in (non-)rotating star clusters of \citet{Webbetal2019}, who use very different initial conditions to the work presented here (Plummer distribution with $5\times 10^4$ stars and rotation is induced by simply giving a fraction of stars some additional rotational velocity following \citet{LyndenBell1960}, which is not physical. Distribution functions from, e.g., \citet{Goodman1983a,LongarettiLagoute1996,EinselSpurzem1999,VarriBertin2012} should be used instead). They find that the orbits of BHs that receive velocity kicks of arbitrary origin decay differently depending on the star cluster rotation. The larger the star cluster rotation, the earlier the orbits of the BHs circularise around the cluster centre due to the gain of angular momentum. As a result, dynamical friction becomes less effective in decaying the orbit. This may happen well before the BHs enter the so-called Brownian regime (e.g. \citet{Chatterjeeetal2002,Lingam2018}), where any systematic orbit decay has stopped and the motion of the BHs is random. Due to the slowed down orbital decay with increasing rotation in the pre-Brownian motion regime, there could be more tidal capture events leading to larger BHBH abundances via three-body scatterings, where a MS star in a BHMS binary is exchanged with another BH \citep{Webbetal2019}. \\
	The $n_{\mathrm{WDWD}}$ evolution mirrors that of $n_{\mathrm{cb}}$, where $n_{\mathrm{cb}}$ is offset from $n_{\mathrm{WDWD}}$ mostly by $n_{\mathrm{BHBH}}$. It is unsurprising that the $n_{\mathrm{cb}}$ is dominated by $n_{\mathrm{WDWD}}$ in the long-term and by $n_{\mathrm{BHBH}}$ in the beginning of simulation, because the massive stars evolve much faster than low mass stars and also our IMF contains many more low mass stars than high mass stars.
	
	\section{Summary, conclusion and perspective}
	\label{Section:Summary, conclusion and perspective}
	\subsection{Summary}
	For the first time we have studied the impact of initial bulk rotation, realistic stellar evolution mass loss models \citep{Kamlahetal2021} in combination with primordial binaries and stars drawn from a continuous IMF \citep{Kroupa2001} as well as a tidal field mass loss on the global dynamics of the star clusters and the development, evolution and coupling of the gravothermal and the gravogyro catastrophes using direct $N$-body methods. We have therefore expanded upon but also greatly surpassed any previous study on this phenomenon in astrophysical realism \citep{EinselSpurzem1999,Kimetal2002,Kimetal2004,Ernstetal2007,Kimetal2008,Fiestasetal2006,FiestasSpurzem2010,Hongetal2013,Wangetal2016,Szolgyenetal2018,Szolgyenetal2019,Szolgyenetal2021,Tiongcoetal2022,Livernoisetal2022}. \\
	In total, we have run eight simulations over 1~Gyr in total, four with stellar evolution (\texttt{SEV} models) and four without stellar evolution (\texttt{noSEV} models). In each subgroup of the two aforementioned groups, every individual model is distributed with a different rotating King model based on \citet{EinselSpurzem1999}. We use one non-rotating model ($\omega_0=0.0$) and three more models with increasing fractions of the initial total star cluster energy being stored in initial bulk rotational energy ($\omega_0=0.6, 1.2, 1.8$). We make the following observations:
	\begin{itemize}
		\item We obtain the same four phases in the early star cluster evolution that were previously observed in direct $N$-body simulations with low particle numbers by \citet{AkiyamaSugimoto1989} for both the runs with and without stellar evolution. Fig.~\ref{Lagrangian}, Fig.~\ref{triaxility} and Fig.~\ref{Angular_momentum_comparison} can be used in combination to deduce the following: we see a phase of violent relaxation that is followed by the gravogyro catastrophe of finite amplitude, where the amplitude depends on the degree of initial bulk rotation (see Fig.~\ref{triaxility}). This gravogyro catastrophe then levels off and angular momentum is transported from the high mass stars (and compact objects) to the lower mass stars (and compact objects) (see Fig.~\ref{Angular_momentum_comparison}). Simultaneously, the system becomes gravothermally unstable and then collapses (see Fig.~\ref{Lagrangian}). This is direct evidence for the coupling of the gravogyro and the gravothermal catastrophes that was first discussed by \citet{Hachisu1979,Hachisu1982} and it is therefore appropriate to coin this process the gravothermal-gravogyro catastrophe. We also directly observe the predicted overall angular momentum loss from the cluster due to the tidal field in all models \citep{AkiyamaSugimoto1989}. \\
		\item The \texttt{SEV}$\omega_01.2$ and \texttt{SEV}$\omega_01.8$ models evolve as follows: The BHs and their progenitor stars, which were distributed axisymmetrically initially, very quickly form a central bar, which rotates, as they transport angular momentum to lower mass stars and compact objects (see Fig.~\ref{triaxility}, Fig.~\ref{Angular_momentum_comparison} and Fig.~\ref{BH_bar}). The bar then becomes an axisymmetric structure over longer time-scales. the outer halo stars (and compact objects) form a more spherical configuration in the long-term, while the stars (and compact objects) in the centre of the cluster form an axisymmetric structure that more slowly becomes spherical over time. \\
		\item The presence of stellar evolution and the tidal field of the star cluster impacts the aforementioned processes in a way that can be deduced mainly from Fig.~\ref{Lagrangian}, Fig.~\ref{triaxility}, Fig.~\ref{Global_sphericity} and Fig.~\ref{Angular_momentum_comparison}. While the early dynamical evolution between the models with and without stellar evolution is similar qualitatively, the gravothermal-gravogyro catastrophe is stronger and happens slightly earlier in the \texttt{noSEV} models (see Fig.~\ref{Global_properties_newtidal}). Most notably, the systems without stellar evolution evolve to similar configurations in the long-term (spherical halo of lower mass stars and compact objects with an axisymmetric centre of higher mass stars and compact objects), but are generally prohibited by doing so due to strong tidal field mass and angular momentum loss (see Fig.~\ref{Global_properties_newtidal}, Fig.~\ref{Angular_momentum_comparison}, Fig.~\ref{Global_sphericity} and Fig.~\ref{escaper_properties}). Instead they exhibit a second and even a third gravogyro collapse and approach a maximally triaxial state in the limit of 1~Gyr. It is an open question if this effect is dampened by larger initial tidal radius (see Fig.~\ref{Global_properties_newtidal}). \\
		\item The \texttt{noSEV}$\omega_01.2$ and \texttt{noSEV}$\omega_01.8$ models also form a bar of the high mass stars that is more concentrated and more triaxial than the bar that forms with stellar evolution due to the lack of stellar evolution mass loss and compact object natal kicks. This bar becomes axisymmetric over time as well, but is also more compact than the counterparts in the \texttt{SEV} models (see Fig.~\ref{BH_bar}). \\
		\item The models \textit{without} stellar evolution reveal that the $M_{\mathrm{mm}}$ mass group (see Fig.~\ref{Angular_momentum_comparison}) appear to replace the increasingly depleting numbers of $M_{\mathrm{hm}}$ objects in the cluster centre and form a subsystem there in the mid- to long-term cluster evolution. This result implies that mass segregation for the $M_{\mathrm{hm}}$ objects has effectively slowed down significantly at that point in simulation time. \\
		\item There is a significant increase in the number of BHBH binaries, $n_{\mathrm{BHBH}}$, present in the \texttt{SEV}$\omega_01.2$ model (see Fig.~\ref{Binary_abundances}). There are also smaller increases in these numbers later on for the \texttt{SEV}$\omega_00.6$ and \texttt{SEV}$\omega_01.8$ models. However, it could also just be statistical fluctuation. This needs to be explored with further simulations and appropriate initial conditions that especially concern the IMF and the binary (orbital) parameters. 
	\end{itemize}

	\subsection{Conclusion}	
	The inclusion of initial bulk rotation in direct $N$-body simulations of star clusters is still unusual, although it has been known for over a century that star clusters even today show significant imprints of rotation, for example, in their shape \citep{PeaseShapley1917,ShapleySawyer1927,Shapley1930,KopalSlouka1936,King1961,FrenkFall1982,Harris1976,Harris1996,Kormendy1985,WhiteShawl1987,Luptonetal1987,ChenChen2010,Bianchinietal2013}. This work therefore provides a bridge between observations and theory of the gravothermal-gravogyro catastrophe and the angular momentum and heat transport within a star cluster to much greater detail than any of the previous studies (see large body of work listed in Sect.~\ref{Section:Introduction}). However, this is just another milestone on the road to unravel the impact of initial bulk rotation on realistic star clusters because many important questions are yet to be answered. \citet{Rizzutoetal2021a,Rizzutoetal2022,ArcaSeddaetal2021} have shown the formation and growth of an IMBH in a star cluster simulated by the same code as used here; so far our initial stellar density have been less than in their models. The question is what effect has rotation as in our models on the number and growth of IMBH in star clusters? This issue has only been briefly mentioned in Sect.~\ref{Section:Binary stars} of this paper and demands more simulations.
	
	\subsection{Perspective on future simulations}
	Reflecting on the discussion and conclusion above, there are several research objectives that require improvements on the simulations presented in this paper:
	\begin{itemize}
		\item Increasing the particle number will yield to better results on all sorts of statistics, but importantly in the context of this paper, the calculation of $b/a$, $c/a$ and $\tau$ would be significantly improved, especially in the innermost spheroids of the star cluster models. \\
		\item Accordingly, increasing the binary fraction will yield more robust results on compact binary fractions and would enable us to make better and less speculative assessments on how initial bulk rotation affects compact binary formation. \\
		\item Increasing the density of the initial star cluster models will enable us to make assessments on the initial stellar merger rates of BH progenitor stars and subsequently IMBH formation. \\
		\item Extending our study to more flattened systems, to have a steady transition from spheroidal to disky systems. Our current initial models are not well-suited for disky systems, but e.g. \citet{Vergaraetal2021} provide suitable disky rotating models. How do the gravothermal and gravogyro catastrophes proceed in disky systems? In this paper we still used the concept of Lagrangian radii, based on spherical systems (except when computing the principal axes $a,b,c$). The latter has been initiated by \citet{TheisSpurzem1999}, it sorts the particles according to their energy in the system, rather than according to their distance (and spherical mass coordinate) from the center,  which means that the system is - in virial equilibrium - approximately subdivided using equipotential surfaces rather than  spherical shells containing certain fractions of total mass. For strongly flattened systems it is necessary to compute quantities like average masses and velocity dispersions in such new spheroidal shells defined by equipotential surfaces. \\
		\item Using a realistic 3-D tidal field, which is possible to be treated with the \textsc{Nbody6++GPU} code version presented here, to study in detail how much angular momentum is carried away by escapers will enable us to assess how tidal shocks through galactic disk passages affect the rotating star cluster. We could then also compare the simulation results to recent cluster observations (e.g. from \cite{Pangetal2021a,Pangetal2022}). 
	\end{itemize}
	We are in the process of tackling some of these issues with direct $N$-body simulations and we expect many exciting results in the future. Among these, a recent work by \citet{FlamminiDottietal2022} is shedding light on the impact of the initial bulk rotation on the ejection properties of free-floating planets and stars in rotating star clusters.
	
	\section*{Acknowledgements}
	We thank the anonymous referee for reading the manuscript carefully and providing many insightful comments, which have sparked discussions that have significantly improved the paper. The authors gratefully acknowledge the Gauss Centre for Supercomputing e.V. for funding this project by providing computing time through the John von Neumann Institute for Computing (NIC) on the GCS Supercomputer JUWELS at Jülich Supercomputing Centre (JSC).  As computing resources we also acknowledge the Silk Road Project GPU systems and support by the computing and network department of NAOC. AWHK is a fellow of the International Max Planck Research School for Astronomy and Cosmic Physics at the University of Heidelberg (IMPRS-HD). We thank Alice Zocchi, Sambaran Banerjee, Jarrod Hurley, Arek Hypki, Long Wang, Kai Wu, Roberto Capuzzo-Dolcetta, Andreas Just, M.B.N. (Thijs) Kouwenhoven for helpful discussions, collaboration, and hospitality during visits. This work was supported by the Volkswagen Foundation under the Trilateral Partnerships grants No. 90411 and 97778. MG and PB were partially supported by the Polish National Science
	Center (NCN) through the grant UMO-2016/23/B/ST9/02732. PB expresses his great thanks for the hospitality of the Nicolaus
	Copernicus Astronomical Centre of Polish Academy of Sciences where
	some part of the work was done. PB acknowledged supported from the Volkswagen Foundation under the special stipend No. 9B870 (2022), 
	from the Science Committee of the Ministry of Education and Science 
    of the Republic of Kazakhstan (Grants No. AP14869395 -- "Triune 
    model of Galactic center dynamical evolution on cosmological time 
    scale"), and from the National Academy of Sciences of Ukraine under the Main Astronomical Observatory GPU computing cluster project No.~13.2021.MM. MAS acknowledges funding from the European Union’s Horizon 2020 research and innovation programme under the Marie Skłodowska-Curie grant agreement No. 101025436 (project GRACE-BH, PI Manuel Arca Sedda). FFD acknowledge the support of the DFG priority program SPP 1992 “Exploring the Diversity of Extrasolar Planets” under project Sp 345/22-1. FFD also acknowledges support from the XJTLU postgraduate research scholarship and Research Development Fund (grant RDF-16-01-16). NN is grateful for funding by the Deutsche Forschungsgemeinschaft (DFG, German Research Foundation) -- Project-ID 138713538 -- SFB 881 (``The Milky Way System'', subproject B08). XYP and SQ are grateful to the grant of National Natural Science Foundation of China, No: 12173029, and the financial support of the Research Development Fund of Xi'an Jiaotong-Liverpool University (RDF-18--02--32). AT appreciates support by Grants-in-Aid for Scientific Research (17H06360, 19K03907) from the Japan Society for the Promotion of Science.
	
	\section*{Data Availability}
	The data from the runs of these simulations and their initial models will be made available upon reasonable request by the corresponding author. The \textsc{Nbody6++GPU} code version that contains also the \texttt{level C} stellar evolution \citep{Kamlahetal2021} is publicly available\footnote{Link to repositories: git service at \\ {\scriptsize \url{https://github.com/kaiwu-astro/Nbody6PPGPU-beijing}} }. The \textsc{McLuster} version that is used in this paper will be made publicly available. A similar version described in \citet{Levequeetal2022} is publicly available\footnote{{\scriptsize \url{https://github.com/agostinolev/mcluster.git}}}. The stellar evolution in that version differs from the implementation used for this paper. The movies of the full simulations, from which the Fig.~\ref{BH_bar} was produced, will be made available upon reasonable request as well and will be uploaded publicly in the future.

	
	
	\bibliographystyle{mnras}       
	\bibliography{main.bib}   
	
	

	
	

	\bsp	
	\label{lastpage}
\end{document}